%% file: highstring.tex
\documentclass[12pt]{article}
\usepackage[a4paper, bindingoffset=0cm, hmargin={2.5cm, 2cm},vmargin={2.5cm, 2.5cm}]{geometry}

\usepackage{relsize}
\usepackage{graphicx}
\usepackage[english]{babel}
\usepackage[utf8]{inputenc}
\usepackage{comment}

\usepackage[T1]{fontenc}
\usepackage{lmodern}
\usepackage[bookmarks]{hyperref}
\usepackage{multirow}
\usepackage{float}

\usepackage{amsmath}
\usepackage{amssymb}
\usepackage{amsthm}
\usepackage{amsfonts}
\usepackage{mathtools}
\usepackage{dsfont}
\usepackage{braket}

\usepackage{pgfplots}
\usepackage{fancyhdr}
\usepackage{appendix}
\usepackage{bbold}
\usepackage{tablefootnote}
\usepackage{float}
\usepackage{amsmath,bm}
\usepackage{enumerate}
\usepackage{pgfplots}
\usepackage{epstopdf}
\usepgfplotslibrary{fillbetween}
\usetikzlibrary{patterns}
\pgfplotsset{compat=1.12}
\usepackage{mathtools}
\usepackage{caption}
\usepackage{subcaption}
\usepackage{cite}
\usepackage{stackrel}

\usepackage{color}

\let\originalleft\left
\let\originalright\right
\renewcommand{\left}{\mathopen{}\mathclose\bgroup\originalleft}
\renewcommand{\right}{\aftergroup\egroup\originalright}

\newcommand{\h}{\mathrel{\phantom{=}}}

\makeatletter
\newsavebox{\@brx}
\newcommand{\llangle}[1][]{\savebox{\@brx}{\(\m@th{#1\langle}\)}%
	\mathopen{\copy\@brx\mkern2mu\kern-0.9\wd\@brx\usebox{\@brx}}}
\newcommand{\rrangle}[1][]{\savebox{\@brx}{\(\m@th{#1\rangle}\)}%
	\mathclose{\copy\@brx\mkern2mu\kern-0.9\wd\@brx\usebox{\@brx}}}
\makeatother

\allowdisplaybreaks
\input stephan

\numberwithin{equation}{section}

\begin{document}
	
	\pagestyle{empty}
	\begin{center}

		{\bf\LARGE\sc
			High Energy String Theory  \\[3mm] and the Celestial Sphere}
		
		\vspace{1.75cm}
		
		{{Xavier Kervyn\textsuperscript{\it a,b} and\ Stephan Stieberger\textsuperscript{\it a}}
			
			\vspace{0.75cm}

			{\it\small
				\textsuperscript{a}Max-Planck Institut f\"ur Physik,\\ Werner--Heisenberg--Institut, 85748 Garching, Germany \\
			}
			\vspace{0.5cm}
			{\it\small
				\textsuperscript{b}Arnold Sommerfeld Center for Theoretical Physics,\\ Ludwigs--Maximilians-Universit\"at, 80333 Munich, Germany
			}
		}
		
	\end{center}
	\vspace{1.5cm}
	
	\begin{center}
		{\bf Abstract}\\
		\end{center}

		\noindent
		We elaborate on a string world--sheet connection to flat space--time holography. More specifically, in the high energy (zero tension) limit of tree--level string scattering  in flat backgrounds the underlying string world--sheets   can be related to the celestial sphere, with the saddle points of the high energy string description representing points on the celestial sphere. We show that this picture continues to hold at all subleading orders in the quantum fluctuations around this classical configuration. As a consequence there is a dual description of the high energy limit of string theory as the large energy  expansion on the celestial sphere organized by (light) higher spin modes.		This approach points to an  intrinsic construction of celestial conformal field theory (CFT)  by relating it to a (free) world--sheet CFT of string theory. We also elaborate on the  high energy representations of tree--level open and closed  string amplitudes and work out their subleading corrections.  Their number theoretic properties and relevance as amplitudes of  tensionless strings are discussed.
				
	\vspace{1cm}
	\begin{flushright}
		{\small MPP--2025--65}
	\end{flushright}
	
	\clearpage
	\pagestyle{plain}
	\tableofcontents
	\break
\section{Introduction}
\def\h{\frac{1}{2}}

Every order in perturbation theory of gravity violates the unitarity bounds by growing powers of energy. Therefore, it is more reasonable to consider gravity in the framework of string theory, where amplitudes show super--soft ultraviolet (UV) behaviour. At very high energies, string scattering amplitudes  fall off exponentially rather than growing rapidly as in quantum field theories. This behaviour is a direct consequence of the underlying 
 two--dimensional ($D\!=\!2$) conformal field theory (CFT) structure on the string world--sheet. 
String theory is equipped with  an intrinsic length scale, the tension $T$
of the fundamental string:
\be\label{Tension}
T=\fc{1}{2\pi \ap}\ .
\ee
In the infinite tension limit ($\ap\ra0$) strings become point-like and string theory is described by
a quantum field theory. This limit has been studied in much detail, in particular at the level of amplitudes. 
The opposite limit, the tensionless one ($\ap\ra\infty$), is of very stringy nature with the appearance of an infinite number of massless higher--spin states and  quantum gravity effects being strongest.
This (hard scattering) limit may be thought as an un--Higgsed phase of string theory with the appearance of infinite
global symmetries mixing all
the oscillator modes and from its study we should learn something about the underlying symmetries of string theory \cite{Gross:1987ar}. Likewise,  string theory may be considered as a broken symmetry phase of a full higher--spin theory. 
The CFT structure on the string world--sheet controls the properties of string amplitudes. It softens their UV behaviour and provides the organizing framework for excited string states. More precisely,  the Virasoro constraints and conformal symmetry regulate the kind of interactions which are allowed. In addition, modular invariance of the CFT dictates how loop amplitudes are constructed and ensures UV finiteness at high energies. Hence, it is the world--sheet CFT, which controls many properties of string theory.

A $D\!=\!2$ CFT also appears in the celestial holography program. The latter  uses conformal correlators for the $D\!=\!4$ scattering amplitudes in (asymptotically) flat spacetimes. The
resulting celestial amplitudes are thus highly constrained by the symmetries of the underlying celestial conformal field theory (CCFT). Eventually, this CCFT should allow oe to compute these amplitudes from first principles just like in string theory. 
However, most of the approaches towards this program have been bottom--up so far, investigating the implications of infrared aspects of gravity and
gauge theories and  exploring the structure of celestial amplitudes. 
Moreover, this aspect of flat space holography does not use the detailed
knowledge of UV completion such as string theory. 
Celestial amplitudes  are obtained through a Mellin transform of 
momentum space amplitudes, where energy is integrated
out. Therefore,  celestial amplitudes in field theory are in general 
divergent in the UV, while in string theory these  energy integrals are finite due to the UV softness of string theory \cite{Stieberger:2018edy}.
Initiated by \cite{Stieberger:2018edy}, only a handful of works within the framework of string theory  exists since then
\cite{Jiang:2021csc,Donnay:2023kvm,Castiblanco:2024hnq,Stieberger:2024shv,Bockisch:2024bia}. 
We should remark, that there are also a few other purely field--theoretical proposals leading to well--defined celestial amplitudes.  e.g. turning on a background field ~\cite{Fan:2022vbz,Casali:2022fro,Fan:2022kpp,deGioia:2022fcn,Gonzo:2022tjm,Banerjee:2023rni,Ball:2023ukj,Crawley:2023brz} or considering Eikonal approximation in celestial CFT, which is dressed by an oscillating phase arising from semi--classical effects \cite{deGioia:2022fcn,Adamo:2024mqn}.

One of the central question in the celestial holography program is:
{\it What is one example of a top-down construction of a 2d celestial dual for a string compactification to 4d?} \cite{Strominger}. For  the understanding of  flat space holography it might be rewarding to look at it from a  string perspective and its underlying conformal field theories. Furthermore, one might ask whether there is any connection between the generic standard CFTs on the string world--sheet and the rather involved CCFT  on the celestial sphere. 
 Indeed, the works \cite{Stieberger:2018edy,Castiblanco:2024hnq} derive a
relation of the latter to the $D\!=\!2$ string world--sheet CFT and  suggest extracting the celestial amplitudes from a world--sheet string computation. 
This correspondence is established in the  high--energy  limit of string theory, which makes direct contact with the string world--sheet, at least classically.
In this limit,  the string amplitudes can be approximated by using saddle point methods \cite{Veneziano:1968yb,Gross:1987kza,Gross:1987ar,Gross:1989ge} and the string world--sheet becomes celestial \cite{Stieberger:2018edy}. The saddle points correspond to classical solutions of the world--sheet theory and its underlying CFT determines these solutions.
On the other hand, a stationary phase expansion  describes the corresponding large energy limit on the celestial sphere. This expansion  not only relies on  the same  saddle--points as the high--energy limit of string theory, 
but  can also be identified with the saddle point expansion of the string amplitude. In this work, we show that this picture continues to hold at all subleading orders in large conformal energy.
As a consequence there is a dual description of the high energy limit of string theory as the large energy  expansion on the celestial sphere organized by (light) higher spin modes.		This approach points to an  intrinsic construction of CCFT  by relating it to a (free) world--sheet CFT of string theory.

The present work is structured as follows. In Section 2 we elaborate on the two alternative  representations of open and closed string amplitudes at small and large inverse string tension $\ap$
referring to low-- and high--energy representations, respectively. For this discussion we shall focus on four--point string tree--level. We discuss their number theoretic properties and the role of single--valued projection in these two pictures.
The subleading corrections  of the high energy representations are explicitly derived
and are related to the saddle point expansion. In Section 3 we first review the celestial string amplitude describing open string tree--level superstring scattering. We then  perform a stationary phase expansion on the celestial sphere and relate it to the saddle point expansion of the high energy open string amplitude. Each subleading order in $\tfrac{1}{\ap^k}$ of the high--energy  string expansion is shown
to Mellin transform to the corresponding terms $\tfrac{1}{\beta^k}$ of the stationary phase expansion of the celestial amplitude.
As a consequence, all subleading orders in the celestial amplitudes can be related through Mellin transforms of the underlying subleading
high--energy string expansion. From this matching we derive a correspondence between the string world--sheet and the celestial sphere and their underlying conformal field theories.
In Section 4 we discuss the relevance of the high--energy  open and closed string amplitudes for amplitudes of tensionless strings.
Finally, in Section 5 we review and discuss our results and give some concluding remarks. In particular, we point out a possible connection of our results to recent results in high energy string scattering 
in Anti-de-Sitter (AdS) background.

\section{High energy representation of  string amplitudes}

String amplitudes are analytic functions of the inverse string tension $\ap$. They can be expanded as power series at different values in the parameter space $\ap$. It is expected that the terms of the power series  are comprised by functions depending on the kinematics invariants, weighted by  certain rational or transcendental numbers of a certain degree. For the following discussion we shall focus on four--point string tree--level, albeit we expect similar results to hold for higher--point amplitudes.
For canonical color ordering the tree--level four open superstring amplitude is given by 
\be\label{open}
\Ac(1,2,3,4)=\fc{\Gamma(1-s)\Gamma(1-u)}{\Gamma(1+t)}\ A_{YM}(1,2,3,4)\ ,
\ee
with the kinematic invariants 
\be
s=\ap(p_1+p_2)^2\ \ ,\ \ t=\ap(p_1-p_3)^2\ \ ,\ \ u=\ap(p_2-p_3)^2\ ,
\ee
subject to $s+t+u=0$, momentum conservation $p_1+p_2=p_3+p_4$  and the Yang--Mills subamplitude $A_{YM}(1,2,3,4)$ encoding all kinematical factors. Furthermore, the tree--level four closed superstring amplitude is given by 
\be\label{closed}
\Mc=\pi\;\fc{su}{t}\ \fc{\Gamma(-s)\Gamma(-u)\Gamma(-t)}{\Gamma(s)\Gamma(u)\Gamma(t)}\ A_{YM}(1,2,3,4)\;\tilde A_{YM}(1,2,3,4)\ .
\ee
Note, that in \req{open} and \req{closed} the combinations of Gamma functions originate from the following real and complex Euler integrals, respectively
\begin{align}
\fc{\Gamma(1-s)\Gamma(1-u)}{\Gamma(1+t)}&=-s\;\int_0^1dx\; x^{-s-1}\ (1-x)^{-u}\ ,\label{Worldo}\\[2mm]
\pi\;\fc{su}{t}\ \fc{\Gamma(-s)\Gamma(-u)\Gamma(-t)}{\Gamma(s)\Gamma(u)\Gamma(t)}&=
-s^2\int_{\IC}d^2z\; |z|^{-2s-2}\ |1-z|^{-2u}\ ,\label{Worldc}
\end{align}
subject to the convergence condition $\Re(s)<0,\Re(u)<1$ and $\Re(s)<0,\Re(u)<1,\Re(t)<0$, respectively. For these conditions, which exclude physical poles, the integrals \req{Worldo} and \req{Worldc} can be evaluated 
and then analytically  continued to the physical regions.

\subsection{Small $\boldmath{\ap}$ representation}

In the small $\ap$ representation for canonical color ordering the open string amplitude \req{open} reads \cite{Schlotterer:2012ny}:
\begin{align}
\Ac_0(1,2,3,4)&=\exp\lf\{\sum_{n=1}^\infty\fc{\zeta(2n)}{(2n)}\lf(s^{2n}+u^{2n}-t^{2n}\ri)\ri\}\nonumber\\
&\times\exp\lf\{\sum_{k=1}^\infty\fc{\zeta(2k+1)}{(2k+1)}\lf(s^{2k+1}+u^{2k+1}+t^{2k+1}\ri)\ri\}\ A_{YM}(1,2,3,4)\ .\label{gauge} 
\end{align}
Likewise, for the closed string amplitude \req{closed} we have:
\be
\Mc_0=\pi\fc{su}{t}\exp\lf\{2\sum_{k=1}^\infty\fc{\zeta(2k+1)}{(2k+1)}\lf(s^{2k+1}+u^{2k+1}+t^{2k+1}\ri)\ri\}\ A_{YM}(1,2,3,4)\;\tilde A_{YM}(1,2,3,4)\ .\label{gravity} 
\ee

\subsection{Large $\boldmath{\ap}$ representation}

String amplitudes show a quite different high energy behaviour than field--theory amplitudes.
In fact, at tree--level their form factors expose some exponential suppression in energy.
In the following we shall use the Stirling formula \cite{Gradshteyn:1702455} 
\be\label{Stirling}
\Gamma(z)=(2\pi)^\h\ e^{(z-\h)\ln z}\ e^{-z}\ \exp\lf\{\sum_{k=1}^\infty \fc{B_{2k}}{2k(2k-1)}\;z^{1-2k}\ri\}\ \ \ ,\ \ \ |{\rm arg}z| <\pi\ ,
\ee
with the even Bernoulli numbers, which can be expressed by zeta values of odd negative arguments:
\be\label{rational}
B_{2k}=\fc{(-1)^{k-1}}{2^{2k-1}}\ \fc{(2k)!}{\pi^{2k}}\ \zeta(2k)=-(2k)\ \zeta(1-2k)\ ,\ k\geq1\ .
\ee
In the sequel we may assume $\Re s,\Re u<0$ and $\Re t>0$ subject to $s+u+t=0$ and apply \req{Stirling} for the Gamma factors in \req{open}. Note, that  these conditions are to be chosen when defining \req{open} and \req{closed} as Euler or complex integrals, respectively. Though this choice parameterizes an unphysical region we may access the other regions by analytic continuation. With this information for $s,u\ra-\infty$ and $t\ra+\infty$ we can write the open string amplitude \req{open} as:
\begin{align}
\Ac_{-\infty}(1,2,3,4)&=(2\pi)^{1/2}\;\lf(\fc{su}{t}\ri)^{1/2}\ e^{-s\ln s-u\ln u-t\ln t}\ (-1)^{-s-u}\nonumber\\
&\times\exp\lf\{\sum_{k=1}^\infty\fc{\zeta(1-2k)}{(2k-1)}\lf(\fc{1}{s^{2k-1}}+\fc{1}{u^{2k-1}}+\fc{1}{t^{2k-1}}\ri)\ri\}\ A_{YM}(1,2,3,4)\ . \label{highgauge}
\end{align}
The expression for the physical region $s\ra+\infty$ and $u,t\ra-\infty$ 
follows from rewriting the open string form factor in \req{open} as \cite{Stieberger:2018edy}
\be
\fc{\Gamma(1-s)\Gamma(1-u)}{\Gamma(1+t)}=u\ \fc{\sin(\pi t)}{\sin(\pi s)}\ \fc{\Gamma(-u)\Gamma(-t)}{\Gamma(s)}
\ee
and by using \req{Stirling} we obtain \cite{Stieberger:2018edy}:
\begin{align}
\Ac_{+\infty}(1,2,3,4)&=(2\pi)^{1/2}\;\fc{\sin(\pi t)}{\sin(\pi s)}\;\lf(\fc{su}{t}\ri)^{1/2}\ e^{-s\ln s-u\ln u-t\ln t}\ (-1)^{-u-t}\nonumber\\
&\times\exp\lf\{\sum_{k=1}^\infty\fc{\zeta(1-2k)}{(2k-1)}\lf(\fc{1}{s^{2k-1}}+\fc{1}{u^{2k-1}}+\fc{1}{t^{2k-1}}\ri)\ri\}\ A_{YM}(1,2,3,4)\ . \label{highgaugea}
\end{align}
Due to the different phases $(-1)^{-s-u}$ in \req{highgauge} and $(-1)^{-u-t}$ in \req{highgaugea}, respectively, the transformation between the two asymptotic expansions is discontinuous, which is known as Stokes phenomenon, cf. also \cite{Mizera:2019vvs}.
On the other hand, the  form factor of the closed string amplitude \req{closed} can be converted into the physical domain by using the relation:
\be
\pi\;\fc{su}{t}\ \fc{\Gamma(-s)\Gamma(-u)\Gamma(-t)}{\Gamma(s)\Gamma(u)\Gamma(t)}=-u^2\ 
\fc{\sin(\pi u)\sin(\pi t)}{\sin(\pi s)}\ \fc{\Gamma(-u)^2\;\Gamma(-t)^2}{\Gamma(s)^2}\ .
\ee
With this the high--energy expansion in the physical region becomes:
\begin{align}
\Mc_{+\infty}&=-2\pi\;\fc{\sin(\pi u)\sin(\pi t)}{\sin(\pi s)}\; \fc{su}{t}\ e^{-2s\ln s-2u\ln u-2t\ln t}\nonumber\\
&\times\exp\lf\{2\sum_{k=1}^\infty\fc{\zeta(1-2k)}{(2k-1)}\lf(\fc{1}{s^{2k-1}}+\fc{1}{u^{2k-1}}+\fc{1}{t^{2k-1}}\ri)\ri\}\ A_{YM}(1,2,3,4)\;\tilde A_{YM}(1,2,3,4)\ .\label{highgrava}
\end{align}
A direct application of the Stirling formula \req{Stirling} at \req{closed} may be justified for complex  Mandelstam variables (with $s+t+u=0$ and $|{\rm arg} (\pm s)|<\pi,|{\rm arg} (\pm t)|<\pi,|{\rm arg} (\pm u)|<\pi$) to give:
\begin{align}
\Mc^c_\infty&=i\pi\;\; \fc{su}{t}\ e^{-2s\ln s-2u\ln u-2t\ln t}\nonumber\\
&\times\exp\lf\{2\sum_{k=1}^\infty\fc{\zeta(1-2k)}{(2k-1)}\lf(\fc{1}{s^{2k-1}}+\fc{1}{u^{2k-1}}+\fc{1}{t^{2k-1}}\ri)\ri\}\ A_{YM}(1,2,3,4)\;\tilde A_{YM}(1,2,3,4)\ .\label{highgrav}
\end{align}
A similar expression  has been suggested in \cite{Aprile:2020luw}.

The high--energy expansions \req{highgauge} and \req{highgrav} can be understood from the underlying string world--sheet as the saddle point approximation around the conformal cross ratio of the four string vertices \cite{Veneziano:1968yb,Gross:1987kza,Gross:1987ar,Gross:1989ge}. For a review we refer to \cite{Lee:2015wwa}. In \req{highgauge} and \req{highgrav} the exponential factor describes the leading classical contribution, while the quantum fluctuations around this classical configuration are given by the exponential factors encoding the 
zeta--function $\zeta(1-2k)$ of odd negative arguments. Since the latter are given by rational numbers there should be some underlying point--particle quantum field theory.
In fact, higher--spin theories, which generalize gravity by including massless fields of arbitrary spin, often feature zeta function regularization in amplitude computations. We refer the reader to Section \ref{STensionless} for further comments.

Note, that computing the saddle point approximation for \req{highgaugea} requires regularizing the underlying open string world--sheet integral \req{Worldo} by shifting $s$ and $u$  by a small imaginary constant $i\eps$ and defining an integration contour (generalized Pochhammer contour) transforming from Euklidian to Lorentzian time \cite{Witten:2013pra}. The latter encircles the points $x=0$ and $x=1$ infinitely many times and leads to infinite many complex saddles in different Riemann sheets whose contributions summed up give \req{highgaugea} \cite{Mizera:2019vvs,Yoda:2024pie}. This contour may also be used to numerically evaluate the integrals \req{Worldo} and \req{Worldc} in the physical domain \cite{Eberhardt:2024twy}.

\subsection{Saddle point expansion on the string world--sheet}

The contribution of the classical solution 
to the high--energy regime of the four--point scattering amplitude \req{open}  can be anticipated from the path integral representation:
\be\label{pathI}
	\Ac\sim\int {\cal D} g\ {\cal D} X\ \exp\lf\{-\fc{1}{4\pi \ap}\int d\sigma^1\;d\sigma^2\sqrt g\;g^{\alpha\beta}\;\partial_\alpha X^\mu\partial_\beta X_\mu\ri\}\ \prod_{i=1}^4 V_o(p_i)\ ,
\ee
with the open string vertex operator:
\be
	V_o(p_i)=\int dz_i\;\sqrt g\; e^{\sqrt{\ap} p_i \cdot X(z_i)}\ \ \ ,\ \ \ p_i^2=0\ .
\ee
At high energies the path integral \req{pathI} is dominated by the classical solution 
\be \label{classicS}
	X_c^\mu(z)= i \sqrt{\ap}\ \left( p_1^\mu\;\ln\lf|1-\fc{z}{z_1}\ri| + p_2^\mu\;\ln\lf|1-\fc{z}{z_2}\ri| - p_3^\mu\;\ln\lf|1-\fc{z}{z_3}\ri| - p_4^\mu\;\ln\lf|1-\fc{z}{z_4}\ri|\right) \ ,
\ee
subject to the condition on the conformal cross ratio of the four vertex operator positions $z_i$ 
\be
	x_0=\fc{(z_1-z_2)(z_3-z_4)}{(z_1-z_3)(z_2-z_4)} =-\fc{s}{t}\ ,
	\label{saddlef}
\ee
with $z_{ij}=z_i-z_j$. Inserting \req{classicS} into \req{pathI} yields the leading classical contribution:
\be\label{classicalA}
	\Ac_c\sim x_0^{-s}\ (1-x_0)^{-u}=(-1)^{-s-u} s^{-s}\;t^{-t}\;u^{-u}\ .
\ee

The quantum corrections to the classical solution \req{classicS} give rise to  corrections of the classical amplitude \req{classicalA}, which can be extracted by performing 
an expansion in $\tfrac{1}{\ap}$ of the Stirling factor of \req{highgauge} yielding terms with zeta values of odd negative argument. Alternatively, we may 
undertake the saddle point approximation of the open string integral \req{Worldo}
\be
F(s,u)=(-s)\;\int_0^1 dx\ x^{-s-1}\;(1-x)^{-u}=\fc{\Gamma(1-s)\Gamma(1-u)}{\Gamma(1+t)}
\ee
describing the open string amplitude \req{open}. We consider the limit $s\ra-\infty$ while keeping the ratio
\be\label{ratio} 
a=-\fc{u}{s}<0
\ee
fixed and apply Laplace method (cf. Appendix \ref{SaddleApp}) to the following integral
\begin{align}
F(s,u)&=(-s)\;\int_0^1 dx\ x^{-s-1}\;(1-x)^{-u}=(-s)\;\int_0^1 \fc{dx}{x}\ e^{-s\lf[\ln x-a\; \ln(1-x)\ri]}\nonumber\\[2mm]
&=\sqrt\fc{2\pi a s}{1-a}\ B^s\ \lf\{\ 1+\sum_{n=1}^\infty\fc{C_{2n}}{A^n}\ \ri\}\ , \label{saddleexp}
\end{align}
with 
\be\label{defB}
A=-s\ \ \ ,\ \ \ B=(1-a)^{1-a}\; (-a)^a\ ,
\ee
and the maximum \req{saddlef} given by 
\be\label{saddle}
x_0=\fc{1}{1-a}\equiv -\fc{s}{t}\ ,
\ee 
with  $f(x)=\tfrac{1}{x},\; g(x)=\ln x-a\ln(1-x)$, 
\begin{align}
C_2&=\fc{1}{12}\;\fc{1-a+a^2}{a(a-1)}\ \ \ ,\ \ \ C_4=\fc{1}{288}\;\fc{(1-a+a^2)^2}{a^2(a-1)^2}\ ,\label{c24}\\[1mm]
C_6&=\frac{139 a^6-417 a^5+402 a^4-109 a^3+402 a^2-417 a+139}{51840\ (1-a)^3 a^3}\ ,\label{c6}\\[1mm]
C_8&=-\frac{\left(a^2-a+1\right) \left(571 a^6-1713 a^5+1698 a^4-541 a^3+1698 a^2-1713 a+571\right)}{2488320\ (a-1)^4 a^4}\ .\label{c8}
\end{align}
We have $g''(x_0)=\fc{(1-a)^3}{a}<0$ for \req{ratio}. 

Note, that the coefficients $C_{2n}$ in \req{c24}--\req{c8} entering \req{saddleexp} simply follow from expanding the Stirling expression \req{highgauge}
\be\label{defBk}
C_{2l}=\lf.\exp\lf\{\sum_{k=1}^\infty\fc{\zeta(1-2k)}{(2k-1)}\fc{B_{2k-1}}{s^{2k-1}}\ri\}\ 
\ri|_{\fc{1}{s^l}}\ ,
\ee
with
\be
B_{2k-1}=\fc{1}{(1-a)^{2k-1}}+\fc{1}{a^{2k-1}}-1
\ee
to the appropriate order in $\tfrac{1}{s^n}$.
In \req{saddleexp} the subleading  corrections appear as power--suppressed terms in $\tfrac{1}{\ap}$. These terms encode the effects of integrating out massive higher spin  states at high energies.

\subsection{Periods and double copy}

By definition, periods  can be expressed as integrals of simple rational functions over rational domains \cite{Kontsevich2001}. As a consequence, since the Bernoulli numbers  are rational numbers, Riemann zeta values $\zeta(1-2k)$ of negative odd arguments are periods  due to \req{rational}. 
Therefore, one common property of  both the small and large $\ap$ representations is the appearance of   periods.

However, for  small $\ap$ in the representations \req{gauge} and \req{gravity} we encounter  different subsets of zeta--values than in the high--energy representations \req{highgauge} and \req{highgrav}.
In addition, for the small $\ap$ representations \req{gauge} and \req{gravity} we have the relation \cite{Stieberger:2013wea,Stieberger:2014hba}
\be\label{svpri}
\Mc_0=\pi\;s\;{\rm sv}\Ac_0(1,2,3,4)\  \cdot\ \tilde A_{YM}(1,2,4,3)\ ,
\ee
with the single--valued projection acting on the motivic version of $\Ac_0$ \cite{Brown:2013gia}:
\be\label{svpr}
{\rm sv}\zeta({2k})=0\ \ \ ,\ \ \ {\rm sv}\zeta(2k+1)=2\;\zeta(2k+1)\ \ ,\ \ k\geq1\ .
\ee
On the other hand, the notion of  single--valued projection ${\rm sv}$ and thus the explicit relation \req{svpr} are lost in the high energy representations \req{highgauge} and \req{highgrav}, unless one states ${\rm sv}\zeta(1-2k)=2\zeta(1-2k)$ in lines of \req{svpr}.  It would be interesting to clarify the explicit definition
of single--valued projection in the full parameter space $\ap$. Note, that based on \cite{Stieberger:2014hba} implicitly the single--valued map is defined at the level of the full integrals 
by the means of single--valued integration \cite{Brown:2018omk,Brown:2019wna}.

Interestingly, for the high--energy representation \req{highgauge} and \req{highgrav}
we anticipate the following simple squaring relation:
\be
\Mc^c_\infty=\fc{i}{2}\ \Ac_{-\infty}(1,2,3,4)\; \cdot\;\tilde \Ac_{-\infty}(1,2,3,4)\ .
\ee
Such a double copy relation is possible in the high--energy regime since both expressions \req{highgauge} and \req{highgrav} encode the same subset of zeta values.
On the other hand, the double copy (KLT) relations \cite{Kawai:1985xq} for  \req{highgauge} (and likewise for \req{highgaugea}) give rise to:
\be\label{Highgrav}
\Mc_{\pm\infty}=-\Ac_{\pm\infty}(1,2,3,4)\ \fc{\sin(\pi s)\sin(\pi u)}{\sin(\pi t)}\ \tilde \Ac_{\pm\infty}(1,2,3,4)\ .
\ee
For \req{highgrava} we can express the trigonometric ratio as follows \cite{Schlotterer:2012ny}
\be
\fc{\sin(\pi u)\sin(\pi t)}{\sin(\pi s)}=\pi\;\fc{tu}{s}\ \exp\lf\{-2\sum_{n=1}^\infty\fc{\zeta(2n)}{(2n)}\lf(u^{2n}+t^{2n}-s^{2n}\ri)\ri\}\ ,
\ee
leading for the physical regime   $s\ra+\infty$ and $u,t\ra-\infty$ to the form:
\begin{align}
\Mc_{+\infty}&=-2\pi^2\;u^2\;  e^{-2s\ln s-2u\ln u-2t\ln t}\ \exp\lf\{-2\sum_{n=1}^\infty\fc{\zeta(2n)}{(2n)}\lf(u^{2n}+t^{2n}-s^{2n}\ri)\ri\}\nonumber\\
&\times\exp\lf\{2\sum_{k=1}^\infty\fc{\zeta(1-2k)}{(2k-1)}\lf(\fc{1}{s^{2k-1}}+\fc{1}{u^{2k-1}}+\fc{1}{t^{2k-1}}\ri)\ri\}\ A_{YM}(1,2,3,4)\;\tilde A_{YM}(1,2,3,4)\ .
\end{align}
Though  this form is problematic since it uses expressions of different convergence radius.
At any rate we conclude that our discussion entails the  single--valued map \req{svpri} to be strictly only in place in the low--energy expansion.

It is interesting to note, that in the low--energy expansions \req{gauge} and \req{gravity} we encounter Riemann zeta values of positive argument (i.e. transcendental numbers), while in the high--energy  expansions \req{highgauge} and \req{highgrav}  zeta values of negative odd arguments  (i.e. rational  numbers) appear.
This observation is very reminiscent of the AdS/CFT dressing phase in the N=4 spin chain between weak and strong coupling expansions \cite{Beisert:2006ez}.  Eventually, the two expressions
\req{gauge} and \req{highgauge} and likewise  \req{gravity} and \req{highgrav} should be related by analytic continuation in the inverse string tension parameter $\ap$. These observations  remind of resurgence properties of  infinite sum representations 
and how the various low energy zeta
function coefficients resumm into the high--energy expansion involving Bernoulli numbers
and vice--versa.  In fact, twisted intersection theory might give another relation between the two regimes $\ap\ra0$ and $\ap\ra\infty$
 as it intertwines the two limits. In this formalism field--theory expressions $\ap\ra0$ are computed from string expressions from the high energy limit $\ap\ra\infty$ by localization at saddle--points (or scattering) equations \cite{Mizera:2019gea}.
 
Finally, another note is that  zeta values of negative odd arguments (Bernoulli numbers $B_n$ with $ n\geq0$) appear in the Baker--Campbell--Hausdorff formula, underlying their relevance in quantum field theory. For non--commuting operators $X$ and $Y$ we have
 \be
 e^X\ e^Y=e^{U(X,Y)}\ ,
 \ee
with:
\begin{align}\label{Zexpr}
U(X,Y)&=X+Y+\h\;[X,Y]+\fc{1}{12}\;[X,[X,Y]]-\fc{1}{12}\;[Y,[X,Y]\\ 
&-\fc{1}{24}\;[Y,[X,[X,Y]]-\fc{1}{720}\;[X,[X,[X,[X,Y]]]]-\fc{1}{720}[Y,[Y,[Y,[Y,X]]]]  +\ldots\ .\nonumber
\end{align}
The expression \req{Zexpr} can be decomposed as
\be\label{Zpart}
U(X,Y)=X+U_1(X,Y)+U_2(X,Y)+\ldots\ ,
\ee
with $U_k(X,Y)$ containing all terms of degree $k$ in $Y$.
In particular, for the term linear in $Y$ we have \cite{Mansour:2015awy}:
\be\label{Zpart}
U_1(X,Y)=\sum_{n\geq 0}\fc{(-1)^n B_n}{n!}\;\underbrace{[X,[X,\ldots,[X}_{n},Y]]\ldots]
\ee
This is very reminiscent from the Drinfeld associator $Z(e_0,e_1)$ generating zeta values of positive arguments
from a series of non--commuting group--like elements $e_0,e_1$. Similar to   \req{Zpart} we have the part of the Drinfeld associator
\be
Z_1(X,Y)=\sum_{n\geq 1}\zeta(n+1)\;\underbrace{[e_0,[e_0,\ldots,[e_0}_{n},e_1]]\ldots]
\ee
as series generating $\zeta(n+1)$ in complete analogy to \req{Zpart}.
	
\section{String world--sheet and celestial sphere}

Celestial amplitudes \cite{Pasterski:2016qvg,Pasterski:2017ylz} have been the
central objects in celestial holography. Scattering amplitudes formulated w.r.t. the standard
momentum eigenstate basis are converted into the boost eigenstate basis making conformal
properties manifest. 
For massless external states, the relation between $D=4$ momentum space and celestial
amplitudes is particularly simple. The points on the celestial sphere are related to the asymptotic directions of light--like momenta of external particles. The massless four--momentum
is parameterized by
\be\label{mom}
p^\mu=\omega q^\mu\ ,\quad {\rm with}~~ q^\mu={1\over 2}\ (1+|z|^2, z+\bar z,-i(z-\bar z),1-|z|^2)\ ,
\ee
with $\omega$ the light--cone energy scale and $z\in\IC$ points on the celestial sphere.
The celestial $n$--point amplitude is defined as Mellin transform w.r.t. the conformal weights $\Delta_k$ of the $D=4$ scattering amplitude $\Ac$ in the momentum eigenstate basis:
\be\label{melt}
\tilde{\cal A}_{\{\Delta_l\}}(\{z_l,\bar z_l\})=\bigg(\prod_{k=1}^{n}
\int_0^\infty\omega_k^{\Delta_k-1}d\omega_k\bigg)\ \delta^{(4)}(\omega_1 q_1+\omega_2 q_2-\sum_{m=3}^n\omega_m q_m)\  {\cal A}(\{\omega_l,z_l,\bar z_l\})\ .
\ee
Thus the energies $\omega_k$ are traded for the conformal dimensions $\Delta_k$.
The  amplitude \req{melt} is defined for all $\Delta_k\in\IC$, although a complete and normalizable basis of conformal primary states is built by the continuous principal continuous series $\Delta_k=1+i\lambda_k$, with $\lambda_k\in\IR$. E.g. the discrete series at $\Delta=1-\IZ_+$ corresponds to an expansion w.r.t. soft modes \cite{Donnay:2018neh}. 
Due to momentum conservation delta--function celestial amplitudes are distributions and have only support on certain patches of the celestial sphere, cf. also Subsection 3.3.

In \cite{Stieberger:2018edy} tree--level four-point string scattering amplitudes are transformed into the correlation functions of primary conformal fields on two-dimensional celestial sphere. 
The celestial string amplitude  is an expansion in the inverse cross--ratio and depends on the energies $\lambda$.
In particular, for large cross--ratio $r\ra\infty$ corresponding to small scattering angle we 
recover the field--theory limit \cite{Stieberger:2018edy} and likewise for $r\ra0$ \cite{Donnay:2023kvm} and\footnote{This limit can be anticipated from \req{IntI}
 by noting $\int_0^1{dz\over z}\ \lf( \ln \fc{z}{1-z} \ri)^{\beta-1}=\int_0^1{dz\over z}\ (\ln z)^{\beta-1}$.} $r\ra1$, respectively. In this section we want to elaborate on an expansion in terms of 
large energies $\sum_i\lambda_i\ra\infty$.

Let us  discuss the type I four open superstring scattering amplitude describing the MHV four--gluon amplitude in its field--theory limit.
The corresponding celestial string amplitude may be written in the following form \cite{Stieberger:2018edy}
\be
\tilde \Ac_{STTH}(\{\lambda_i\})=(2\pi)^{-1}\;{\ap}^\beta\ \tilde\Ac_{FT}'(\{\lambda_i\})\ 
a^{-\fc{\beta}{3}}\;(1-a)^{-\fc{\beta}{3}}\ I(a,\beta)\ ,\label{celstring}
\ee
with the celestial gluon amplitude $\tilde\Ac_{FT}(\{\lambda_i\})$
\begin{align}\label{celgluon}
&\qquad\quad\tilde\Ac_{FT}(\{\lambda_i\})=\tilde\Ac_{FT}'(\{\lambda_i\})\ \delta\bigg(\sum_{n=1}^4\lambda_n\bigg)\ ,\\[2mm]
\tilde\Ac_{FT}'(\{\lambda_i\})&=8\pi \ \delta(r-\bar r)\ \Bigg(\prod_{i<j}^4 z_{ij}^{{h\over 3}-h_i-h_j}
\bar z_{ij}^{{\bar h\over 3}-\bar h_i-\bar h_j}\Bigg)\, a^{-\fc{7}{3}}\,(1-a)^{2\over 3}\, \theta(r-1)\ ,
\end{align}
and the conformal cross--ratio 
\be\label{Cross}
r=\frac{1}{a}=\fc{z_{12}z_{34}}{z_{23}z_{41}}
\ee 
as well as  $h=\sum_{n=1}^4h_n$ and $\bar h=\sum_{n=1}^4\bar h_n$. The kinematic invariants $t,u$ can be  expressed in terms of \req{Cross} as $u=-sa$ and $t=s(a-1)$. Up to some numerical factors, the expression \req{celgluon} is in agreement   with \cite{Pasterski:2017ylz}. Furthermore, we have the parameter:
\be\label{beta}
\beta:= -{i\over 2}\sum_{k=1}^4\lambda_k\equiv -\h\sum_{k=1}^4(\Delta_k-1)\ .
\ee
In \req{celgluon} the cross ratio $r$ is constrained  to be real by the delta--function $\delta(r-\bar r)$, which follows from momentum conservation. It means that the four points $z_i$ are required to lie on a circle on the celestial sphere.
In addition, only the celestial field--theory amplitude \req{celgluon} requires delta--function support on  $\sum_{i=1}^4\lambda_i$, i.e. involves the factor $\delta(\beta)$. 
The celestial amplitude \req{celgluon} has the conformal transformation properties of a four--point correlation function of primary conformal fields with weights
\begin{align}
&h_1={i\over 2}\lambda_1,\qquad\qquad \bar h_1=1+{i\over 2}\lambda_1,\cr
&h_2={i\over 2}\lambda_2,\qquad\qquad \bar h_2=1+{i\over 2}\lambda_2,\cr
&h_3=1+{i\over 2}\lambda_3,\qquad ~\bar h_3={i\over 2}\lambda_3\ ,\cr
&h_4=1+{i\over 2}\lambda_4,\qquad ~\bar h_4={i\over 2}\lambda_4\ ,\label{weight}
\end{align}
in agreement with $\Delta_n=1+i\lambda_n$, $J_1=J_2=-1$ and $J_3=J_4=+1$.

The celestial string amplitude \req{celstring} follows from performing four Mellin transforms w.r.t. energies $\lambda_i$ of the open string amplitude \req{open}.
The genuine string part of \req{celstring} is encoded in the function \cite{Stieberger:2018edy}
\begin{align}
I(a,\beta)&=-\fc{a^\beta}{2}\int_0^\infty dw\; w^{-\beta-1}\ F\lf(\tfrac{w}{a},-w\ri)\label{IntIa}\\[2mm]
&=-\h\; \Gamma(1-\beta)\ \int_0^1{dz\over z}\ \lf[\ \ln z-a\ln(1-z)\ \ri]^{\beta-1}\ ,\label{IntI}
\end{align}
encapsulating all string modes.
In \cite{Stieberger:2018edy} this integral has been evaluated  by expressing it as a power series in the inverse cross ratio $r=\tfrac{1}{a}$ dressed with some periods. In the limit $a\ra0$ the celestial four--gluon amplitude \req{celgluon} is recovered. Due to $PSL(2,\IR)$ invariance, this limit is equivalent to $a\ra\infty$. 
We have already stressed above, that the celestial string amplitude is not restricted to $\beta=0$, but $\beta$ is an arbitrary complex parameter, i.e.  $\beta\in\IC$.
In this work we want to perform for generic cross ratio \req{Cross} an expansion w.r.t. large $\beta$. 

\subsection{Stationary phase expansion  on the celestial sphere}

In this subsection we want to apply the method of stationary phase approximation on the integral \req{IntI} to investigate the limit $\beta\ra\pm i\infty$.
For this we apply results from Appendix \ref{SaddleApp} with: 
\begin{align}
g(x)&=\ln\lf[\;\ln x-a\ln(1-x)\;\ri]\ ,\\
f(x)&=\fc{1}{x\;\lf[\ln x-a\ln(1-x)\ri]}\ .
\end{align}
The function $g$ has a stationary point at \req{saddle}, with 
\be
g^{(2)}(x_0)= \fc{(a-1)^3}{a \ln B} \neq0\ , \label{g(x)}
\ee
and $B$ defined in \req{defB}.
 After gathering all terms we arrive at:
\begin{align}
I(a,\beta)&=\Gamma(1-\beta)\ \ a^{\tfrac{1}{2}}\;(1-a)^{-\tfrac{1}{2}}\ \sqrt{\fc{2\pi}{\beta}}\ (\ln B)^{\beta-\tfrac{1}{2}}\label{expansionI}\\
&\times \Big\{1+\fc{1}{\beta}\lf(\fc{1}{8}-C_2\ln B\ri)+\fc{1}{\beta^2}\lf(\fc{1}{128}+\fc{3}{8}\ C_2\ln B+C_4\ln^2B\ri)\cr
&+\fc{1}{\beta^3}\lf(-\fc{5}{1024}-\fc{25}{128}\ C_2\ln B-\fc{15}{8}\ C_4 \ln^2B-C_6\ln^3 B\ri)\cr
&+\fc{1}{\beta^4}\lf(-\fc{21}{32768}+\fc{105}{1024}\ C_2\ln B+\fc{385}{128} C_4\ \ln^2B+\fc{35}{8}C_6\ln^3 B+C_8\ln^4B\ri)+\ldots\Big\}\ .\nonumber
\end{align}
The first and leading term in \req{expansionI} has already been determined in \cite{Stieberger:2018edy}. One interesting observation is, that the stationary phase expansion \req{expansionI} in $\tfrac{1}{\beta}$ is determined by the 
expressions \req{defBk} stemming from the Stirling factor of \req{highgauge}. The latter describes the subleading terms of the saddle point approximation of the string form factor \req{saddleexp} with coefficients $C_{2l}$ given by \req{defBk}.
Actually, for $\Re\beta<0$ we can rearrange the terms in \req{expansionI} to arrive at:
\begin{align}
I(a,\beta)&=\sqrt{2\pi}\ \Gamma\lf(\frac{1}{2}-\beta\ri)\ a^{\tfrac{1}{2}}\;(1-a)^{-\tfrac{1}{2}}\ (\ln B)^{\beta-\tfrac{1}{2}}\label{expansionIa}\\[1mm]
&\times \Bigg\{1-\fc{C_2\;\ln B}{\beta+\tfrac{1}{2}}+\fc{C_4\;\ln^2 B}{\lf(\beta+\tfrac{1}{2}\ri)\lf(\beta+\tfrac{3}{2}\ri)}-\fc{C_6\;\ln^3 B}{\lf(\beta+\tfrac{1}{2}\ri)\lf(\beta+\tfrac{3}{2}\ri)\lf(\beta+\tfrac{5}{2}\ri)}+\ldots\Bigg\}\ .\nonumber
\end{align}
In this form we have a one--to--one correspondence between subleading orders in $\tfrac{C_{2l}}{\beta^l}$ and the subleading orders in $\fc{C_{2l}}{\ap^l}$ of the string form factor \req{saddleexp}.
In fact, the result \req{expansionIa} follows directly  from performing at each order in  $\fc{1}{\ap}$  of the saddle point expansion of the  open string amplitude \req{saddleexp} four Mellin transforms w.r.t. the energies $\lambda_i$. These transformations boil down to computing (for $s,a<0$) the final integral:
\begin{align}\label{todo}
\int_{-\infty}^0ds\; s^{-\beta-1}\;\sqrt\fc{2\pi a s}{1-a}\ B^s\ \fc{C_{2n}}{(-s)^n}&=
(-1)^n\;C_{2n}\ \sqrt\fc{2\pi a }{1-a}\int_{-\infty}^0ds\; s^{-\beta-\tfrac{1}{2}-n}\;\ B^s\ .
\end{align}
One may  rewrite the integral above as:
\begin{align}
	\int_{-\infty}^0ds\; s^{-\beta-\tfrac{1}{2}-n}\;\ B^s &= (\ln B)^{\beta - \frac{1}{2} + n} \int_{-\infty}^0 dt \, t^{- \left(\beta + \frac{1}{2} + n \right)} e^t \cr
	&= (\ln B)^{\beta - \frac{1}{2} + n} \int_{0}^\infty dt \, (-t)^{- \left(\beta + \frac{1}{2} + n \right)} e^{-t}\ .
\end{align} 
To proceed, we use the Hankel representation for the reciprocal of the Gamma function \cite{Whittaker_Watson_1996}
\be
	\fc{1}{\Gamma(z)} = \frac{i}{2\pi}\ \lim_{\delta \to 0^+} \int_{C_\delta} dt \; (-t)^{-z}\; e^{-t}, \quad z\in \IC\backslash\IZ\ ,
\ee where the Hankel contour $C_\delta$ is a path on the Riemann sphere from $+\infty$ inbound along the real line to $\delta > 0$, counterclockwise around a circle of radius $\delta$ at 0, back to $\delta$ on the real line, and outbound
back to $+ \infty$ along the real line. The contour $C$ does not cross $[0,\infty)$ .In other words, for $z = \beta + \frac{1}{2} + n$ we have
 \begin{align}
	- \frac{2\pi i}{\Gamma(z)} &= \lim_{\delta \to 0^+} \left\{ \int_{\infty}^\delta dt \, (- t)^{-z} e^{-t} + \text{(integral over small circle)} + \int_\delta^\infty dt \, e^{- 2\pi i z} (-t)^{-z} e^{-t} \right\}\nonumber\\
	&= (e^{-2 \pi i z} - 1) \int_0^\infty dt \, (-t)^{-z} e^{-t} = - 2 i e^{-\pi i z} \sin(\pi z) \int_0^\infty dt \, (-t)^{-z} e^{-t}\ ,
\end{align}  
where we kept track of the appropriate analytic continuation of $(-t)^{-z}$ around its branch cut along the positive real axis, given that $\beta \in i \IR$. Eventually, we find
\be
	\int_0^\infty dt \, (-t)^{-\left( \beta + \fc{1}{2} + n \right)}\ e^{-t} = \fc{\pi\; (-1)^{\beta + \fc{1}{2}}}{\Gamma\left( \beta + \fc{1}{2} + n \right)}\ \cos(\pi \beta)^{-1}\ ,
\ee 
and thus \req{todo} becomes:
\be
\int_{-\infty}^0ds\; s^{-\beta-1}\;\sqrt\fc{2\pi a s}{1-a}\ B^s\ \fc{C_{2n}}{(-s)^n} = (-1)^n\;C_{2n} \cos(\pi \beta)^{-1} \sqrt\fc{2\pi a }{1-a} \fc{\pi (-1)^{\beta + \fc{1}{2}} }{\Gamma\left( \beta + \fc{1}{2} + n \right)} (\ln B)^{\beta - \frac{1}{2} + n}\ .\label{Corres}
\ee
Hence,  with $\tfrac{\pi}{\cos(\pi\beta)}=\Gamma(\tfrac{1}{2}-\beta)\Gamma(\tfrac{1}{2}+\beta)$ we arrive at the expression \req{expansionIa}.

It is important to note, that both the expansion
\req{expansionIa} and  series \req{saddleexp} are around the same saddle point \req{saddle} and likewise \req{saddlef} expressed by celestial coordinates $z_i$ in \req{Cross}.
As a consequence both  the open superstring 
amplitude \req{open} and the celestial string amplitude \req{celstring} localize on the same 
point on the string world--sheet and celestial sphere, respectively. Additionally, both saddle--point approximations are in one--to--one correspondence to all subleading orders in the order $\tfrac{1}{\ap}$ and  $\tfrac{1}{\beta}$, respectively. Furthermore, let us give also an alternative expression for \req{expansionIa}
\begin{align}
I(a,\beta)&=\sqrt{2\pi}\ \Gamma\lf(\frac{1}{2}-\beta\ri)\ a^{\tfrac{1}{2}}\;(1-a)^{-\tfrac{1}{2}}\ (\ln B)^{\beta-\tfrac{1}{2}}\nonumber\\[1mm]
&\times \Bigg\{\ 1-\fc{C_2\;\ln B-C_4\;\ln^2 B+\tfrac{1}{2}C_6\;\ln^3 B+\ldots}{\beta+\tfrac{1}{2}}-
\fc{C_4\;\ln^2 B-C_6\;\ln^3 B+\ldots}{\beta+\tfrac{3}{2}}\nonumber\\[1mm]
&-\fc{\tfrac{1}{2}C_6\;\ln^3 B+\ldots}{\beta+\tfrac{5}{2}}+\ldots\ \Bigg\}\ ,\label{expansionIaa}
\end{align}
which explicitly exposes the residua at $\beta=-\tfrac{1}{2},-\tfrac{3}{2},\ldots$.

An other note concerns  the high--energy string representation \req{saddleexp} entering \req{todo}.
To discuss the asymptotic large energy behaviour $\Re\beta\ra\pm\infty$ of \req{expansionI} and \req{expansionIa} the detail of the analytic continuation of \req{highgauge} to the physical regime
\req{highgaugea} does not matter in the expression \req{IntI}.

\subsection{String soft modes and UV completion}

Since celestial amplitudes follow from momentum space by Mellin transforms w.r.t. energies they are sensitive to both UV and IR. This feature translates into
powerful constraints on the analytic structure of celestial amplitudes.  
Low- and high--energy features of celestial massless four--point amplitudes are characterized by 
their analytic structure in the complex $\beta$--space, with $\beta$ expressed in \req{beta} in terms of the four energies $\lambda_i$. 
In quantum field theory celestial amplitudes are meromorphic functions in the complex $\beta$ space and exhibit simple poles at all  integers $\beta\in\IZ$. There are residua along the positive axis from IR physics and poles in the UV region along the negative axis \cite{Arkani-Hamed:2020gyp}. The latter are expected to  generically disappear in a complete theory of quantum gravity like string theory, where amplitudes show an exponential fall off, cf. \req{highgauge} and \req{highgrav}. 
Note, that in the celestial string amplitude \req{celstring} the only dependence on the string scale appears as universal factor $\ap^\beta$ as a result of   integrating over all of the energies in the amplitude. As a consequence the usual hierarchy among the 
low--energy and the high--energy regimes is blurred.

The expansion \req{expansionI} represents a  power series expansion in the inverse parameter $\beta$. 
Though we have derived it from the limit $\beta\ra\pm i\infty$ by performing a stationary phase approximation, the representation for $\beta\ra\pm\infty$ follows analogously\footnote{Strictly speaking, the representation for $\beta\ra+\infty$ can be straightforwardly obtained by applying Laplace approximation, since the saddle point is then at a maximum of $g$ in \eqref{g(x)}. The same approximation with $\beta \to - \infty$ for $\beta < 0$ would instead correspond to a minimum of $g$ and fall outside  the context of validity of Laplace approximation. We thus hereby assume that this limit is approached from the complex plane and not on the real axis, such that the stationary phase approximation may still be used.}, cf. e.g. \req{expansionIa}.
Soft factorization of celestial amplitudes is realized in Mellin space exhibiting conformally soft particles. The Gamma factor $\Gamma(1-\beta)$ furnishes residua at all positive integers $\beta\in\IZ_+$ due to IR effects. This factor stems from the Mellin representation of  \req{IntI}
\begin{align}
\int^\infty_0dw\; w^{-\beta-1}\; w\; e^{-w X}&=\Gamma(1-\beta) \ X^{\beta-1}\nonumber\\
&=\fc{1}{1-\beta}-\fc{X}{2-\beta}+\fc{\tfrac{1}{2}X^2}{3-\beta}+\ldots\ ,
\end{align}
with $X=\ln x-a\ln(1-x)$ from \req{IntI} and $\Re X>0$. The tower of residua corresponds to the usual celestial soft--modes in the IR.
Since for the latter we have $\Delta =1+i\lambda\in 1-\IZ_+$ we may interpret these states as soft modes of the high energy string expansion \req{expansionI}.

On the other hand, in the limit  $\Re\beta\ra-\infty$ the Gamma factor $\Gamma(1-\beta)$ exhibits the expected black hole dominance. This is in agreement with the fact, that we are considering an UV completed theory provided by string theory.  By applying
\be
\Gamma(z)\sim(2\pi)^{\frac{1}{2}}\; \; z^{z-\frac{1}{2}}\; e^{-z}\ \times
\begin{cases} 1\ , & \Re z>0\ ,\\
\fc{1}{e^{2\pi i z}-1}\ ,&\Re z<0\ ,
\end{cases}
\ee
following from \req{Stirling} and $\Gamma(1-\beta)=(-\beta)\;\Gamma(-\beta)$ we can  expand the first line of \req{expansionI} w.r.t. large beta $\Re\beta\ra\pm\infty$, 
\begin{align}\label{stemm}
\Gamma(1-\beta)\ & a^{\tfrac{1}{2}}\;(1-a)^{-\tfrac{1}{2}}\ \sqrt{\fc{2\pi}{\beta}}\ (\ln B)^{\beta-\tfrac{1}{2}}\\
                 &=-2\pi i\ \lf(\fc{-\beta}{e}\ri)^{-\beta}\ \ a^{\tfrac{1}{2}}\;(1-a)^{-\tfrac{1}{2}}\ (\ln B)^{\beta-\tfrac{1}{2}}
                   \times\begin{cases} 1\ , & \Re\beta\ra-\infty\ ,\\
                                                                                    \fc{1}{e^{-2\pi i \beta}-1}\ ,&\Re\beta\ra+\infty\ ,  \end{cases}\nonumber
\end{align}
and experience an  exponential growth for $\Re\beta\ra-\infty$. For  $\Re\beta\ra+\infty$ we observe the factor $\tfrac{1}{e^{-2\pi i \beta}-1}$, which
  accounts for the infinite tower of poles $\beta\in\IZ_+$ in IR and is probed by the lightest string threshold  \cite{Arkani-Hamed:2020gyp}. This factor is universal for all string  mass levels $\sqrt{n/\ap}, n\in\IZ_+ $ which appear as residua from massive string poles in the form \cite{Stieberger:2018edy}:
\be
{I}(a,\beta)
=\pi\,\delta(\beta)+{i\pi\over (1-e^{-2\pi i\beta})}\ \sum_{n=1}^\infty{\rm Res}_{s=n}\bigg\{s^{-\beta}\! B\Big( {-}s,1+{s\over r}\Big)\bigg\}\ .
\ee  
Thus, the region at $\Re\beta\ra+\infty$  is controlled by the exchange of massive string levels $n$ in the low--energy string expansion. 

\begin{figure}[H]
    \centering
    \includegraphics[scale=.4]{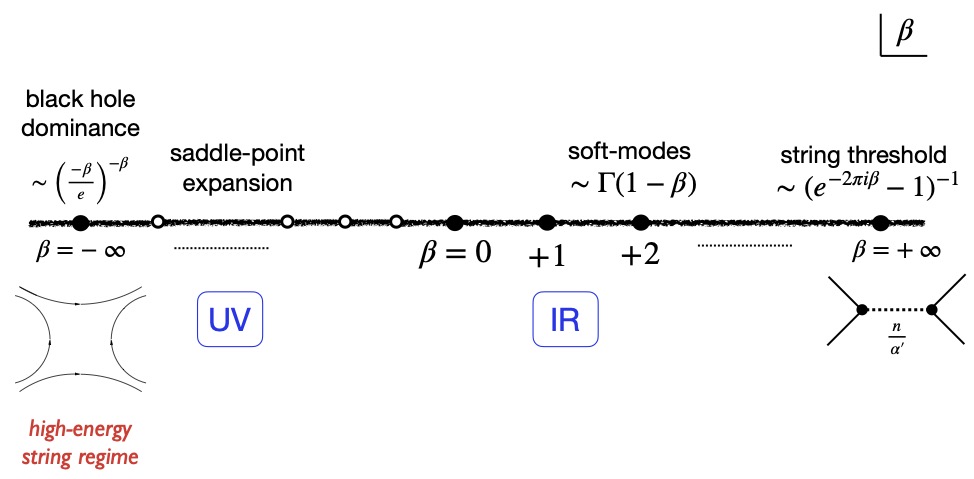}
    \caption{Complex $\beta$ plane with UV and IR regions and corresponding string threshold.}
    \label{BetaFig}
\end{figure}

For negative real $\beta$ the integral \req	{IntI} converges and is analytic in $\beta$ and   no poles along the negative $\beta$--axis are expected from \req{expansionI}. 
However, the series \req{expansionIa} and \req{expansionIaa} expose poles for negative real $\beta$, where the light higher spin states of the high--energy string theory become relevant. They give rise to poles 
$\beta=-\tfrac{1}{2},-\tfrac{3}{2},\ldots$ along the negative axis and can directly be attributed to the subleading string corrections of the high--energy expansion \req{saddleexp}. While the string saddle--point approximation \req{saddleexp} yields leading exponential falloff the subleading  corrections appear as power--suppressed terms in $\tfrac{1}{\ap}$. These corrections include both terms associated with higher--spin exchanges (string resonances) and local contact terms organized as a derivative expansion. In the $\beta$--plane \req{beta} these effects correspond to operators with $\Delta_k\geq2$  and manifest as  poles along the negative axis.  
In the expansion \req{expansionIa}   these  subleading corrections   in $\tfrac{1}{\beta^k}$ with residua up to $\ln^kB$ account for those higher string modes. From the Mellin integral \req{Corres} we formally have the correspondence between string 
($\tfrac{1}{\ap}$) and celestial ($\tfrac{1}{\beta}$) expansion:
 \begin{align}
B^{\ap}\ \fc{C_{2k}}{\ap^k}\  &\stackrel[Eq.\; \req{Corres}]{\mbox{Mellin}}{\longleftrightarrow}\ \ap^\beta\; (-1)^k\;C_{2k}\ \fc{(\ln B)^{\beta - \frac{1}{2} + k}}{\cos(\pi\beta)\;\Gamma\left( \beta + \fc{1}{2} + k \right)}
\nonumber\\
&\sim  \ap^\beta\;(-1)^k\;C_{2k}\ \fc{(\ln B)^{\beta - \frac{1}{2} + k}}{(\beta+\frac{1}{2} )\cdot\ldots\cdot(\beta-\frac{1}{2} + k)}\ .
\end{align}
These poles in $\tfrac{1}{\beta^k}$ are in one-to-one correspondence with the subleading corrections in $\tfrac{1}{\ap^k}$ of the high--energy string expansion
  \req{saddleexp} describing quantum fluctuation around the classical string solution \req{classicS}
  and accounting for light higher--spin modes. The latter correspond to the string spectrum in the 
ultra high--energy regime  of string theory. 
According to \req{defBk}  these corrections are tied to  combinations of zeta functions of odd negative argument $\zeta(1-2l)$ constituting transcendentality degree $k$.
To summarize our discussion in Fig. \ref{BetaFig} we display the $\beta$--space and its properties from UV to IR. 
 
\noindent 

\noindent
Eventually, the poles along the negative $\beta$--axis are superposed in the limit $\Re\beta\ra-\infty$ by the leading power  behaviour $\sim\lf(\tfrac{\frac{1}{2}-\beta}{e}\ri)^{\frac{1}{2}-\beta}$ of \req{expansionIa} which is also anticipated   in \req{stemm}. Interestingly, in the large UV limit $\Re\beta\ra-\infty$ we detect a similar behaviour than in the Eikonal approximation  \cite{Adamo:2024mqn}. Such a regulating phase has also been suggested for the UV region in  \cite{Arkani-Hamed:2020gyp}.
In our case this limit describes the high--energy super--Planckian limit, where the string world--sheet and celestial sphere are related.
A thoroughly discussion of the analytic structure in the full complex $\beta$--space and likewise expansions in different regimes of $\ap$  is in progress 
with  powerful  relations between different regimes to be expected.

\subsection{String world--sheet as celestial sphere}

The open string moduli space (at tree--level)  is given by the configuration space that parameterizes isomorphism classes of $n$ ordered points along the boundary of the disk (which is a real projective line) modulo M\"obius transformations $PSL(2,\IR)$
\begin{equation}\label{Space}
\left[({\bf RP}^1)^n-\bigcup_{i,j,k,l}\left\{r_{ijkl}=0,1\right\}\right]/PSL(2,\IR)\ ,\ n\geq4\ ,
\end{equation}
with the $n-3$ algebraically independent conformal cross--ratios $r_{ijkl}=\tfrac{z_{ij}z_{kl}}{z_{ik}z_{jl}}$, cf. e.g. \cite{Stieberger:2016xhs}. Note, that the $PSL(2,\IR)$ group allows to fix three punctures at $0,1$ and $\infty$ along the boundary of the disk (or complex upper half--plane).
Points $x_i$ on the space \req{Space} correspond to open string vertex operator positions
of the underlying CFT. The complexified version of \req{Space} is the moduli space 
$\mathcal{M}_{0,n}$ of $n$ marked points on the Riemann sphere up to M\"obius transformations relevant to (tree--level) closed string insertions on the world--sheet sphere and their dynamics:
\be\label{SpaceCC}
\mathcal{M}_{0,n}
=\left[({\bf CP}^1)^n-\bigcup_{i,j,k,l}\left\{r_{ijkl}=0,1\right\}\right]/PSL(2,\IC)\ ,\ n\geq4\ .
\ee
On the other hand, points $z_i$ on the celestial sphere 
are associated to the asymptotic directions of (light-like) momenta $p_j^i$ of external particles as:
\be\label{CPoints}
 z_j=\frac{p^1_j+ip^2_j}{p^0_j+p^3_j}=\frac{p^0_j-p^3_j}{p^1_j-ip^2_j}\ ,\ \ \ j=1,\ldots,n\ .
 \ee
 The moduli space for the generic $n$-point celestial correlators is \cite{Mizera:2022sln}
\be\label{cSpace}
\left[({\bf CP}^1)^n-\bigcup_{i,j,k,l}\left\{\Im(r_{ijkl})=0\right\}\right]/PSL(2,\IC)\ ,\ n\geq5\ ,
\ee
which appears to be a non--linear variant  of \req{SpaceCC}. Generically, for $n\geq5$ the $n-3$ cross--ratios are complex numbers.

For the case $n=4$ discussed in this work due to $SL(2,\IC)$ invariance the only non--trivial dependence on the four points $z_i,\; i=1,\ldots,4$ is through the cross--ratio \req{Cross}. Momentum conservation in $D=4$ requires the spatial momenta for all four particles to lie on a two--dimensional plane, which intersects the
celestial sphere on a circle. Consequently, the four points $z_i$ on the celestial sphere must be located on this circle, which translates into the condition $\Im r=0$. Note, that this reality condition is reminiscent of the moduli space \req{Space} of open string vertex operators to be inserted on the boundary of a disk and a direct map between \req{saddle} and \req{Cross} 
is furnished. 

The saddle point approximation of the string amplitude \req{open} makes closest contact to the underlying string world--sheet \req{Space}. Likewise,  the stationary phase expansion \req{expansionI} of the celestial string amplitude \req{celstring} provides the most direct contact to the  celestial sphere. In \cite{Stieberger:2018edy} we have raised the question whether there is any limit in which vertex operator positions are tied to the celestial sphere and we have evidenced that the analog of the high--energy ultra--Planckian limit $\ap\ra\infty$ is reached at $\lambda\ra\infty$. In this limit the string world-sheet becomes celestial or likewise the celestial sphere becomes the string  world--sheet. Note, that this map is possible since the delta--function constraint $\delta(r-\ov r)$ of the celestial string amplitude \req{celstring} is supported by the  (open string) saddle point \req{saddle}. In other words, the amplitude description  of open strings on the disk world--sheet provides automatically this reality constraint.
The series \req{expansionIa} represents a stationary phase expansion of the string form factor \req{IntI} on the celestial sphere with the stationary point \req{saddle} expressed in terms of the celestial coordinates $z_i$ and the kinematic cross--ratio  in \req{Cross}.
This identifies a point  on the string world--sheet \req{Space} with a point \req{CPoints} on the celestial sphere  -- up to $SL(2,\IR)$ transformations. As a consequence, in this limit the underlying  celestial CFT computing  the celestial string form factor \req{IntI} and the free CFT determining the string form factor \req{open} should become identical. Likewise, the infinite--dimensional  Virasoro
symmetry  on the open  string world--sheet should relate to the symmetries on the celestial sphere. Again,  this identification becomes possible since the delta--function constraint $\delta(r-\ov r)$ of the celestial string amplitude \req{celstring} is supported by the (string) saddle point solution \req{saddle}.

Hence, by embedding our celestial gluon amplitude \req{celgluon} into string theory we 
can write this amplitude as a saddle--point approximation with the (string) saddle point \req{saddle} localized on the celestial sphere and the subleading quantum corrections in $\tfrac{1}{\beta}$ specified by light higher spin modes from the string spectrum in the ultra--Planckian limit $\ap\ra\infty$.
Our high-energy relation between string world--sheet and celestial sphere may establish 
an underlying  vertex operator description of CCFT, which in particular reproduces the known celestial OPEs.  Furthermore, this approach may lead to an  intrinsic construction of the CCFT by relating it to a (free) world--sheet CFT of string theory\footnote{More precisely,  the underlying world--sheet symmetry of the (open)  closed tensionless string is a (boundary) Carrollian conformal symmetry ($\partial\mbox{CCS}_2$) $\mbox{CCS}_2$, cf. the next section. Hence, the latter symmetries should play a role on the celestial sphere describing the (open) closed celestial string amplitude.}.

Generalization of our results to higher point amplitudes $n\geq5$ and matching \req{Space} and \req{cSpace} would be very interesting.
The celestial string amplitude involving five open strings has been computed in \cite{Castiblanco:2024hnq}. In the latter  work also the high--energy string limit has been matched with the large energy limit on the celestial sphere. 
However note, that for higher--point ($n\geq5$) string amplitudes  their high--energy  limit is 
constituted by a sum over $(n-3)!$ (independent) saddle point solutions  following from solving the saddle--point equations  or scattering equations \cite{Gross:1987kza,Gross:1987ar,Gross:1989ge}
\be\label{SQE}
\sum_{j=1\atop j\neq i}^n\fc{p_i\;p_j}{z_i-z_j}=0\ \ \ ,\ \ \ i=1,\ldots,n\ ,
\ee
and specifying  the $n-3$ cross ratios $r_{ijkl}$. They give rise to  a set of $n-3$ independent non--linear equations. The same set of  $(n-3)!$  saddle--point
solutions  arises in the stationary phase approximation of the corresponding celestial $n$--point amplitude in the large energy limit. Through \req{SQE} each of these sets of $n-3$ points (or cross--ratios) is expressed  in terms of the kinematic invariants $p_ip_j$, lies on the space \req{SpaceCC} (or \req{Space} for appropriately chosen kinematics) and is to be matched with points \req{CPoints} of the configuration space  \req{cSpace}. Some of these solutions may give more dominant contributions to the saddle--point expansion than others. However, for each scattering process  there is only one single kinematical configuration \req{CPoints} of $n$ points on the celestial sphere \req{cSpace}, which does fulfill \req{SQE}, but  represents only one of  $(n-3)!$ solutions of \req{SQE}. 
Actually, there is a special kinematic configuration (Eikonal constraints) for which the equations \req{SQE} are solved by one single solution \cite{Dvali:2014ila}. The latter then maps to the single  kinematical configuration \req{CPoints} on the celestial sphere. 
At any rate, there remains more to be clarified for $n\geq5$ how to define for a generic kinematic configuration a bijektive mapping of all of the $(n-3)!$ solutions of \req{SQE} onto \req{CPoints} and \req{cSpace}.

\section{Comments on amplitudes of tensionless strings}	\label{STensionless}

In this section we want to discuss the high--energy representations $\Ac_\infty,\Mc_\infty$ of the gauge \req{highgauge} and gravitational \req{highgrav} amplitudes.
In the limit $\ap\ra\infty$ the string masses become small, leading to an infinite tower of nearly massless states of mass $\sqrt{n/\ap}, n\in\IZ_+$ contributing to $\Ac_\infty,\Mc_\infty$. Their dynamics resembles a system with an infinite number of massless higher--spin (of arbitrary integer number) states contributing similarly to the amplitudes \req{highgauge} and \req{highgrav} and a higher--spin symmetry is exhibited.
Thus, the limit $\ap\ra\infty$ is expected to probe the behaviour of string theories at very high energies in similar way as the field theory limit $\ap\ra0$ probes the short--distance  properties of ordinary point--particle theories. 
However, in the limit $\ap\ra\infty$ the effective field theory description breaks down due to the infinite number of massless states  contributing to $\Ac_\infty,\Mc_\infty$.

The high--energy limit $\ap\ra\infty$ is related\footnote{The high--energy limit $\ap\ra\infty$ is  inequivalent to the null string proposed by Schild as 
 tensionless limit of the Nambu--Goto string  at fixed sigma--model field coordinates
rather than at fixed oscillator variables \cite{Schild:1976vq}.} to the tensionless limit of open and closed string theory, where the usual relativistic string picture breaks down. No techniques for straightforwardly computing amplitudes in this regime are available so far.
Classically, the null string is the tensionless limit of the usual string theory sigma--model \cite{Schild:1976vq}. 
The tensionless string (albeit fundamental on its own) can be related to the tensile one in two limits of coordinates $\sigma,\tau$ on the world--sheet  \cite{Bagchi:2015nca}.
More specifically, in the non--relativistic limit the length of the string becomes infinite (i.e. $\sigma\ra\infty$), with the world--sheet coordinates
\be\label{tenslimitinf}
(\sigma,\tau)\lra\lf(\fc{1}{c}\;\sigma,\tau\ri)\ \ \ ,\ \ \ c\lra\infty\ ,
\ee
and the (world-sheet) speed of light $c$ going to infinity. Secondly, in the  ultra--relativistic or Carrollian limit on the world--sheet with  the (world-sheet) speed of light $c$ going to zero \cite{Bagchi:2013bga}:
\be
(\sigma,\tau)\lra(\sigma,c\;\tau)\ \ \ ,\ \ \ c\lra0\ .
\ee 
The mode expansion of the closed tensionless string can be obtained from the mode expansion of the tensile closed string
\be
X^\mu(\sigma,\tau)=x^\mu+2\sqrt{2\alpha'}\;\alpha_0^\mu\;\tau+i\sqrt{2\alpha'}\ \sum_{n\neq0}\frac{1}{n}\ \left\{\ \alpha_n^\mu\;e^{-in(\tau+\sigma)}+\tilde \alpha_n^\mu\; e^{-in(\tau-\sigma)}\ \right\} 
\ee 
 by taking the limit 
 \be\label{tenslimit}
 {\sigma\rightarrow \sigma\atop \tau\rightarrow c\tau}\ \ ,\ \ \alpha'\rightarrow\frac{1}{c}\ ,\ c\rightarrow0\ ,
 \ee
 yielding
 \be
 X^\mu(\sigma,\tau)=x^\mu+\sqrt{2}\;A_0^\mu\;\sigma+\sqrt{2}\;B_0^\mu\;\tau+i\sqrt{2}\sum_{n\neq0}\frac{1}{n}\;(A_n^\mu-in\tau\;B_n^\mu)\; e^{-in\sigma}\ ,
 \ee
with the new oscillators for the tensionless string:
\begin{align}
A_n^\mu&=\frac{1}{\sqrt c}\;(\alpha_n^\mu-\tilde\alpha_{-n}^\mu)\ ,\\
B_n^\mu&=\sqrt c\;(\alpha_n^\mu+\tilde\alpha_{-n}^\mu)\ .
\end{align}
A similar limit \req{tenslimit}
can be performed on the open tensile string  to yield the open null string  with Dirichlet boundary conditions \cite{Bagchi:2024qsb}. 

By applying   the limits \req{tenslimit} the subleading corrections in $\tfrac{1}{\ap}$ of  \req{highgrav} may describe and map to corrections in $c$ of the amplitudes involving closed tensionless strings, i.e.:
\be
\frac{1}{\ap}\sim c\ .
\ee
Likewise, the subleading corrections in $\tfrac{1}{\ap}$ of the open tensile string amplitude \req{highgauge} should describe corrections in $c$ of  amplitudes involving open null strings with Dirichlet boundary conditions.

The distinction between open and closed string spectra is strictly only in place for the tensile strings \cite{Francia:2007qt}. On the other hand, the distinction between closed and open strings becomes blurred in the tensionless limit of string theory \cite{Bagchi:2019cay} suggesting a phase where strings become effectively massless and more like tensionless flux tubes.
As a consequence an open and closed string and their  world--sheets  look similar in this limit. This explains why at tree--level the open string amplitude \req{highgauge} and the closed string amplitude \req{highgrav} assume qualitatively the same analytic forms with the same  subset of zeta values. Furthermore, since only rational numbers appear in their expansions, a field--theory description (notably a higher--spin gauge theory) should be accessible.

The conformal symmetry on the open tensile string world--sheet is an infinite--dimensional  Virasoro
symmetry ${\rm Vir}$, while for the tensile closed string a double copy of the latter ${\rm Vir}\otimes\overline{\rm Vir}$ is exhibited. These infinite--dimensional algebras and  the properties of  the underlying two--dimensional conformal field theory on the string--world sheets are crucial to compute in the usual tensile string theories amplitude results like \req{gauge} and \req{gravity}.
On the other hand, in the tensionless limit \req{tenslimitinf} of the closed string a conformal 
Galilean $\mbox{CGal}_2$ world--sheet symmetry  emerges. Likewise, in the tensionless limit \req{tenslimit} a Carrollian  $\mbox{CCS}_2$ world--sheet symmetry  in two dimensions emerges   \cite{Bagchi:2013bga}.  The latter is isomorphic\footnote{Carrollian CFTs are
field theories exhibiting conformal Carroll (or BMS) symmetries as spacetime
symmetries, and can be constructed from standard CFTs by formally taking
the speed of light to zero. Carrollian amplitudes behave as correlation 
functions of Carrollian CFT that holographically describes asymptotically flat
spacetime.} to the $D\!=\!3$ Bondi--Metzner--Sachs ($\mbox{BMS}_3$) algebra appearing in the context of asymptotically flat spacetimes at their null boundary. Furthermore, recently 
a boundary Carrollian CFTs $\partial\mbox{CCS}_2$ has been constructed for the null limit  of the tensile open string emerging by contracting
a single copy of the Virasoro algebra \cite{Bagchi:2024qsb}. It is believed that these symmetries play a similar role for the tensionless string than the Virasoro symmetry for the tensile string.
While the closed string amplitude \req{gravity} is computed by  methods explicitly using the holomorphic/anti--holomorphic factorization ${\rm Vir}\otimes\overline{\rm Vir}$ on the string world--sheet (by means of KLT method) \cite{Kawai:1985xq} it would 
be rewarding to understand the corresponding analog of KLT and the role of 	$\mbox{CCS}_2$ in the world--sheet derivation of \req{highgrav}.

\section{Concluding remarks}
	
\begin{figure}[H]
    \centering
    \includegraphics[scale=.45]{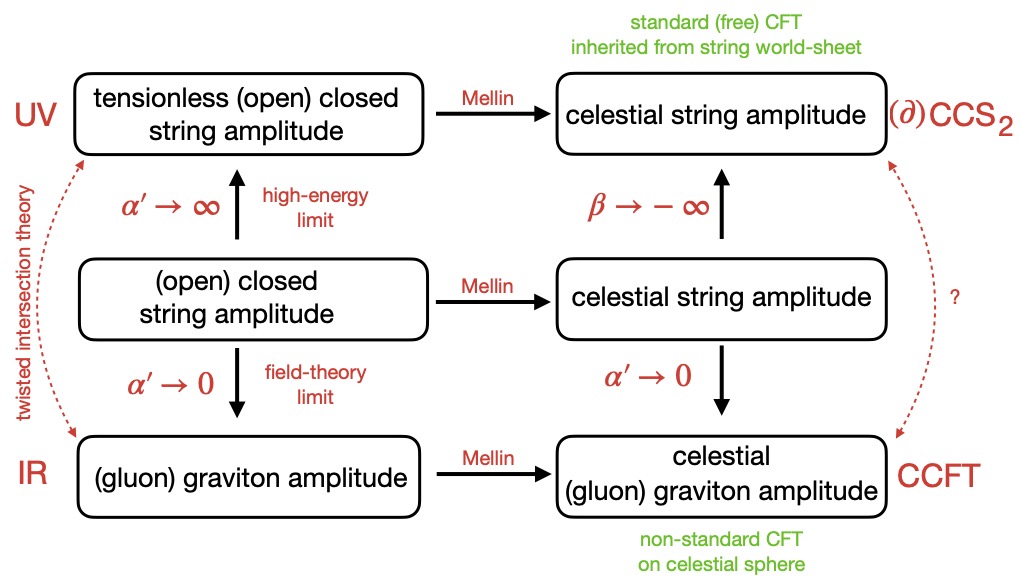}
    \caption{String amplitude vs. celestial amplitude for $\ap\ra0$ and $\ap\ra\infty$.}
    \label{KLTtreeFig}
\end{figure}

In Section 2 we have presented explicit expressions \req{highgauge} and \req{highgrav} for the open and closed string amplitudes in the high--energy (zero string tension $\ap\ra\infty$) limit  at tree--level in flat backgrounds, respectively. In particular, we have worked out their subleading corrections and analyzed their number theoretic structure described by Bernoulli numbers or zeta values with negative arguments. As a consequence we evidenced 
that the explicit definition of
single--valued projection \req{svpr} is strictly well--defined only  for expansions w.r.t. small $\ap\ra0$  and needs to be redefined for the opposite limit $\ap\ra\infty$.
In Section 3  we have established a correspondence between the celestial open string amplitude \req{celstring} and the open string amplitude \req{open} in the zero tension limit. The saddle point expansion of the high energy string amplitude \req{saddleexp} directly relates to the stationary phase expansion \req{expansionI} of the celestial string amplitude \req{IntI}. In \req{Corres} this map can be established order by order in the subleading corrections  $\tfrac{1}{\ap}$ and $\tfrac{1}{\beta}$, respectively. Thus our results generalize the observation made in \cite{Stieberger:2018edy} to all orders. In Figure \ref{KLTtreeFig} we show the infinite and zero tension limits of the closed (open) string amplitude, their corresponding celestial counterparts following after Mellin transformation
and their interrelations by taking certain limits.
\noindent 
We have argued that the formulation of  asymptotically flat space--times through a CCFT on the celestial sphere may have a string realization through a standard CFT on string world--sheet.
The Mellin transformation of the high--energy limit of the closed (open) string amplitude should bequeath its underlying  world--sheet CFT -- some (boundary) Carrollian $\mbox{CCS}_2$ -- to the celestial string amplitude in the limit $\beta\ra-\infty$. 
 We believe, that studying the UV limits $\ap\ra\infty$ and likewise $\beta\ra-\infty$  could help to understand the underlying CCFT in the IR. While there is a relation between the field--theory $\ap\ra0$ and high--energy $\ap\ra\infty$ limits of string theory, via twisted intersection theory (likewise CHY formalism) it is not clear how this relation translates after performing Mellin transformation.
At any rate embedding celestial holography into string theory should teach us how to make profit from the underlying standard CFT on the string world--sheet.

It would be interesting to generalize our results to AdS and 
find  the corresponding celestial dual of high energy string scattering in AdS background (string theory on $AdS\times S^5$).
More specifically, recently  the four--graviton amplitude in AdS has been proposed in the high--energy 
limit as \cite{Alday:2023pzu}
\begin{align}
  \Mc_{AdS}&=e^{-2s\ln s-2u\ln u-2t\ln t}\ W_0(x_0)\ \exp\lf\{\ \fc{s^2}{R^2}\;V_3(x_0)+\fc{s^3}{R^4}\;V_5(x_0)+\ldots\ \ri\}\nonumber\\
  &=e^{-2s\ln s-2u\ln u-2t\ln t}\ W_0(x_0)\ \lf(1+\fc{s^2}{R^2}\;W_3(x_0)+\fc{s^4}{R^4}\;W_6(x_0)+\ldots\ri) \ ,\label{AdSgrav}
\end{align}
with a rational function $W_0$, some transcendental functions $V_i,W_i$ and large $R,s$ and $s^2/R^2$ fixed, where $R$ is the common radius of AdS and $S^5$.
This form, which is supposed to also arise from a world--sheet picture in the small curvature expansion \cite{Alday:2023jdk}, is very reminiscent of \req{highgrav} and its corresponding saddle point expansion in terms of zeta values with odd negative weights. In fact, both \req{highgrav} and 
\req{AdSgrav} localize on the same saddle point \req{saddle}. Therefore,  we speculate that the subleading corrections of \req{highgrav} encapsulated by the series of odd negative zeta values should resemble the expansion \req{AdSgrav} and its subleading corrections w.r.t. the inverse radius $\tfrac{1}{R}$.

	\ \\

	\ \\
{\bf Acknowledgements:}
We are grateful to Sebastian Mizera for pointing us to reference \cite{Aprile:2020luw}. In addition, we thank Nima Arkani-Hamed, Arjun Bagchi, Johannes Broedel, Tomasz Taylor and Bin Zhu for both illuminating and valuable discussions.
This work is supported by the DFG grant 508889767 {\it 
Forschungsgruppe ``Modern foundations of scattering amplitudes''}. XK gratefully acknowledges the hospitality of the Niels Bohr Institute in Copenhagen, Denmark, where part of this research was conducted, as well as the generous support of the Fondation du Domaine de Villette.

\vskip3cm

\break
\appendix

\section{Asymptotic expansions of integrals}\label{SaddleApp}

In this section, we introduce the Laplace method, steepest descent method, and stationary phase method, which are key asymptotic techniques for approximating integrals of the form 
\be
	F(A) \equiv \int_C dz \, f(z)\;  e^{A g(z)}
\ee
for large $A > 0$ and $C$  an integration contour in the complex $z$--plane  \cite{miller_applied_2006}. The Laplace method applies when $g$ is real and has a global maximum at $z_0$, where the integral is approximated using a Gaussian expansion around $z_0$. The stationary phase method, on the other hand, is used for highly oscillatory integrals where $g$ is pure imaginary, and contributions come from stationary points where $g'(z) = 0$; thus capturing oscillatory behaviour with an additional phase shift. Finally, the steepest descent method generalizes these ideas to complex functions by deforming the integration contour $C$ above to pass through \textit{saddle points}, where $g'(z)=0$ and the integral decays most rapidly. We now specialize our discussion to the Laplace and stationary phase methods below.

\subsection{Laplace method}

The saddle point method is an extension of the original method of
Laplace for approximating certain integrals.
Laplace's method can be used for obtaining the asymptotic behaviour of integrals in which the large parameter $A$ appears in an exponential \cite{Bender1979AdvancedMM}. Then the major contributions to the value of the integral arise from neighbourhoods of the points at which the exponential attains its greatest value. The latter may be a maximum or supremum. 
For two real continuous functions $f,g$  this  approximation  $A \ra+\infty$ yields
\be
\int_p^q dx\ f(x)\ e^{Ag(x)}=f(x_0)\ e^{Ag(x_0)}\ \sqrt{\fc{2\pi}{-A g''(x_0)}}\ \ 
\lf\{\ 1+\sum_{n=1}^\infty\fc{C_{2n}}{A^n}\ \ri\}\ ,
\ee
with the maximum $x_0$ (with $g'(x_0)=0$ and $g^{(2)}(x_0)<0$) strictly occurring in the region $p<x_0<q$ and \cite{Schafer}
\begin{align}
C_2&=\frac{{f_1} {g_3}}{2 {f_0} {g_2}^2}-\frac{{f_2}}{2 {f_0} {g_2}}-\frac{5 {g_3}^2}{24 {g_2}^3}+\frac{{g_4}}{8 {g_2}^2}\ ,\\[2mm]
C_4&=-\frac{35 {f_1} {g_3}^3}{48 {f_0} {g_2}^5}+\frac{35 {f_1} {g_3} {g_4}}{48 {f_0} {g_2}^4}-\frac{{f_1} {g_5}}{8 {f_0} {g_2}^3}+\frac{35
   {f_2} {g_3}^2}{48 {f_0} {g_2}^4}-\frac{5 {f_2} {g_4}}{16 {f_0} {g_2}^3}-\frac{5 {f_3} {g_3}}{12 {f_0} {g_2}^3}+\frac{{f_4}}{8
   {f_0} {g_2}^2}+\frac{385 {g_3}^4}{1152 {g_2}^6}\nonumber\\[2mm]
   &-\frac{35 {g_3}^2 {g_4}}{64 {g_2}^5}+\frac{7 {g_3} {g_5}}{48 {g_2}^4}+\frac{35 {g_4}^2}{384
   {g_2}^4}-\frac{{g_6}}{48 {g_2}^3}\ ,\\[2mm]
C_6&=\frac{5005 {f_1} {g_3}^5}{2304 {f_0} {g_2}^8}-\frac{5005 {f_1} {g_3}^3 {g_4}}{1152 {f_0} {g_2}^7}+\frac{77 {f_1} {g_3}^2 {g_5}}{64 {f_0}
   {g_2}^6}+\frac{385 {f_1} {g_3} {g_4}^2}{256 {f_0} {g_2}^6}-\frac{7 {f_1} {g_3} {g_6}}{32 {f_0} {g_2}^5}-\frac{21 {f_1} {g_4}
   {g_5}}{64 {f_0} {g_2}^5}+\frac{{f_1} {g_7}}{48 {f_0} {g_2}^4}\nonumber\\[2mm]
   &-\frac{5005 {f_2} {g_3}^4}{2304 {f_0} {g_2}^7}+\frac{385 {f_2} {g_3}^2
   {g_4}}{128 {f_0} {g_2}^6}-\frac{21 {f_2} {g_3} {g_5}}{32 {f_0} {g_2}^5}-\frac{105 {f_2} {g_4}^2}{256 {f_0} {g_2}^5}+\frac{7 {f_2}
   {g_6}}{96 {f_0} {g_2}^4}+\frac{385 {f_3} {g_3}^3}{288 {f_0} {g_2}^6}-\frac{35 {f_3} {g_3} {g_4}}{32 {f_0} {g_2}^5}\nonumber\\[2mm]
   &+\frac{7 {f_3}
   {g_5}}{48 {f_0} {g_2}^4}-\frac{35 {f_4} {g_3}^2}{64 {f_0} {g_2}^5}+\frac{35 {f_4} {g_4}}{192 {f_0} {g_2}^4}+\frac{7 {f_5} {g_3}}{48
   {f_0} {g_2}^4}-\frac{{f_6}}{48 {f_0} {g_2}^3}-\frac{85085 {g_3}^6}{82944 {g_2}^9}+\frac{25025 {g_3}^4 {g_4}}{9216 {g_2}^8}\nonumber\\[2mm]
   & -\frac{1001 {g_3}^3
   {g_5}}{1152 {g_2}^7}-\frac{5005 {g_3}^2 {g_4}^2}{3072 {g_2}^7}
   +\frac{77 {g_3}^2 {g_6}}{384 {g_2}^6}+\frac{77 {g_3} {g_4} {g_5}}{128
   {g_2}^6}+\frac{385 {g_4}^3}{3072 {g_2}^6}-\frac{{g_3} {g_7}}{32 {g_2}^5}-\frac{7 {g_4} {g_6}}{128 {g_2}^5}\nonumber\\[2mm]
   &-\frac{21 {g_5}^2}{640
   {g_2}^5}+\frac{{g_8}}{384 {g_2}^4}\ ,\\[2mm]
 C_8&=\frac{37182145 {g_3}^8}{7962624 {g_2}^{12}}-\frac{1616615 {f_1} {g_3}^7}{165888 {f_0} {g_2}^{11}}-\frac{11316305 {g_4} {g_3}^6}{663552
   {g_2}^{11}}+\frac{1616615 {f_2} {g_3}^6}{165888 {f_0} {g_2}^{10}}+\frac{1616615 {f_1} {g_4} {g_3}^5}{55296 {f_0} {g_2}^{10}}\nonumber\\[2mm]
   &+\frac{323323 {g_5}
   {g_3}^5}{55296 {g_2}^{10}}-\frac{85085 {f_3} {g_3}^5}{13824 {f_0} {g_2}^9}+\frac{8083075 {g_4}^2 {g_3}^4}{442368 {g_2}^{10}}-\frac{425425 {f_2}
   {g_4} {g_3}^4}{18432 {f_0} {g_2}^9}-\frac{85085 {f_1} {g_5} {g_3}^4}{9216 {f_0} {g_2}^9}\nonumber\\[2mm]
   &-\frac{85085 {g_6} {g_3}^4}{55296 {g_2}^9}+\frac{25025
   {f_4} {g_3}^4}{9216 {f_0} {g_2}^8}-\frac{425425 {f_1} {g_4}^2 {g_3}^3}{18432 {f_0} {g_2}^9}+\frac{25025 {f_3} {g_4} {g_3}^3}{2304 {f_0}
   {g_2}^8}-\frac{85085 {g_4} {g_5} {g_3}^3}{9216 {g_2}^9}\nonumber\\[2mm]
   &+\frac{5005 {f_2} {g_5} {g_3}^3}{768 {f_0} {g_2}^8}+\frac{5005 {f_1} {g_6}
   {g_3}^3}{2304 {f_0} {g_2}^8}+\frac{715 {g_7} {g_3}^3}{2304 {g_2}^8}-\frac{1001 {f_5} {g_3}^3}{1152 {f_0} {g_2}^7}-\frac{425425 {g_4}^3
   {g_3}^2}{73728 {g_2}^9}+\frac{25025 {f_2} {g_4}^2 {g_3}^2}{2048 {f_0} {g_2}^8}\nonumber\\[2mm]
   &+\frac{1001 {g_5}^2 {g_3}^2}{1024 {g_2}^8}-\frac{5005 {f_4}
   {g_4} {g_3}^2}{1536 {f_0} {g_2}^7}+\frac{5005 {f_1} {g_4} {g_5} {g_3}^2}{512 {f_0} {g_2}^8}-\frac{1001 {f_3} {g_5} {g_3}^2}{384 {f_0}
   {g_2}^7}+\frac{5005 {g_4} {g_6} {g_3}^2}{3072 {g_2}^8}-\frac{1001 {f_2} {g_6} {g_3}^2}{768 {f_0} {g_2}^7}\nonumber\\[2mm]
   &-\frac{143 {f_1} {g_7}
   {g_3}^2}{384 {f_0} {g_2}^7}-\frac{143 {g_8} {g_3}^2}{3072 {g_2}^7}+\frac{77 {f_6} {g_3}^2}{384 {f_0} {g_2}^6}+\frac{25025 {f_1} {g_4}^3
   {g_3}}{6144 {f_0} {g_2}^8}-\frac{5005 {f_3} {g_4}^2 {g_3}}{1536 {f_0} {g_2}^7}-\frac{1001 {f_1} {g_5}^2 {g_3}}{1280 {f_0}
   {g_2}^7}\nonumber\\[2mm]
   &+\frac{77 {f_5} {g_4} {g_3}}{128 {f_0} {g_2}^6}+\frac{5005 {g_4}^2 {g_5} {g_3}}{2048 {g_2}^8}-\frac{1001 {f_2} {g_4} {g_5}
   {g_3}}{256 {f_0} {g_2}^7}+\frac{77 {f_4} {g_5} {g_3}}{128 {f_0} {g_2}^6}-\frac{1001 {f_1} {g_4} {g_6} {g_3}}{768 {f_0}
   {g_2}^7}-\frac{1001 {g_5} {g_6} {g_3}}{3840 {g_2}^7}\nonumber\\[2mm]
   &+\frac{77 {f_3} {g_6} {g_3}}{192 {f_0} {g_2}^6}-\frac{143 {g_4} {g_7} {g_3}}{768
   {g_2}^7}+\frac{11 {f_2} {g_7} {g_3}}{64 {f_0} {g_2}^6}+\frac{11 {f_1} {g_8} {g_3}}{256 {f_0} {g_2}^6}+\frac{11 {g_9} {g_3}}{2304
   {g_2}^6}-\frac{{f_7} {g_3}}{32 {f_0} {g_2}^5}+\frac{25025 {g_4}^4}{98304 {g_2}^8}\nonumber\\[2mm]
   &-\frac{5005 {f_2} {g_4}^3}{6144 {f_0} {g_2}^7}+\frac{385
   {f_4} {g_4}^2}{1024 {f_0} {g_2}^6}-\frac{1001 {g_4} {g_5}^2}{5120 {g_2}^7}+\frac{231 {f_2} {g_5}^2}{1280 {f_0} {g_2}^6}+\frac{77 {g_6}^2}{7680
   {g_2}^6}-\frac{7 {f_6} {g_4}}{128 {f_0} {g_2}^5}-\frac{1001 {f_1} {g_4}^2 {g_5}}{1024 {f_0} {g_2}^7}\nonumber\\[2mm]
   &+\frac{77 {f_3} {g_4} {g_5}}{128
   {f_0} {g_2}^6}-\frac{21 {f_5} {g_5}}{320 {f_0} {g_2}^5}-\frac{1001 {g_4}^2 {g_6}}{6144 {g_2}^7}+\frac{77 {f_2} {g_4} {g_6}}{256 {f_0}
   {g_2}^6}+\frac{77 {f_1} {g_5} {g_6}}{640 {f_0} {g_2}^6}-\frac{7 {f_4} {g_6}}{128 {f_0} {g_2}^5}+\frac{11 {f_1} {g_4} {g_7}}{128 {f_0}
   {g_2}^6}\nonumber\\[2mm]
   &+\frac{11 {g_5} {g_7}}{640 {g_2}^6}-\frac{{f_3} {g_7}}{32 {f_0} {g_2}^5}+\frac{11 {g_4} {g_8}}{1024 {g_2}^6}-\frac{3 {f_2}
   {g_8}}{256 {f_0} {g_2}^5}-\frac{{f_1} {g_9}}{384 {f_0} {g_2}^5}+\frac{{f_8}}{384 {f_0} {g_2}^4}-\frac{{g_{10}}}{3840 {g_2}^5}\ ,
     \end{align}
   with  $f_i=f^{(i)}(x_0),\; g_i=g^{(i)}(x_0)$.

\subsection{Method of stationary phase}

A different situation arises when the function $g$ has a  complex part or likewise $A$ is imaginary
as 
 \be
\int_p^q dx\ f(x)\ e^{iAg(x)}\ ,
\ee
with two real continuous functions $f,g$.
Then for large $A$ the integrand oscillates vary rapidly due to the phase factor $e^{iAg(x)}$. 
As a consequence the contributions to the integral cancel except near a stationary point where the oscillation is less rapid since the phase is stationary. According to Stokes and Kelvin the dominant contribution to the integral comes from  the region around $g'(x_0)=0$ and we can proceed as in the Laplace case.
For large $A\ra\infty$ we obtain
\be
\int_p^q dx\ f(x)\ e^{iAg(x)}=\begin{cases}
f(x_0)\ e^{iAg(x_0)-\tfrac{1}{4}\pi i}\ \sqrt{\fc{2\pi}{-A g''(x_0)}}\ \ 
\lf\{\ 1+\sum\limits_{n=1}^\infty\fc{C_{2n}}{(-iA)^n}\ \ri\}\ ,&g^{(2)}(x_0)<0\ ,\\[4mm]
f(x_0)\ e^{iAg(x_0)+\tfrac{1}{4}\pi i}\ \sqrt{\fc{2\pi}{A g''(x_0)}}\ \ 
\lf\{\ 1+\sum\limits_{n=1}^\infty\fc{C_{2n}}{(-iA)^n}\ \ri\}\ ,&g^{(2)}(x_0)>0\ ,
\end{cases}
\ee
with the stationary point  $x_0$ (with $g'(x_0)=0$) strictly occurring in the region $p<x_0<q$
and the coefficients $C_{2n}$ given in the previous section  \cite{Mirkov}.

	\newpage
	\bibliographystyle{mystyle}
	\bibliography{./ref}
	\markboth{Bibliography}{Bibliography}

\end{document}

%% file: stephan
\newcommand{\req}[1]{(\ref{#1})}

\def\fc#1#2{\frac{#1}{#2}}

\newcommand{\nwc}{\newcommand}
\nwc{\ba}  {\begin{array}}
\nwc{\ea}  {\end{array}}
\nwc{\bdm} {\begin{displaymath}}
\nwc{\edm} {\end{displaymath}}

\nwc{\bea} {\begin{equation}\ba{lcl}}
\nwc{\eea} {\ea\end{equation}}

\nwc{\be} {\begin{equation}}
\nwc{\ee} {\end{equation}}

\nwc{\bda} {\bdm\ba{lcl}}
\nwc{\eda} {\ea\edm}

\nwc{\bc}  {\begin{center}}
\nwc{\ec}  {\end{center}}

\nwc{\ds}  {\displaystyle}

\nwc{\nn} {\nonumber}
\nwc{\nnn} {\nonumber \vspace{.2cm} \\ }
\nwc{\ra}{\rightarrow}
\nwc{\lra}{\longrightarrow}
\def\lf{\left}\def\ri{\right}

\nwc{\p} {\partial}
\def\IR{{\bf R}}

\def\ap{\alpha'}

\def\Mc{{\cal M}}

\def\ov{\overline}

\def\eps{\epsilon}

\def\IR{{\bf R}}
\def\IC{{\bf C}}

\def\IZ{{\bf Z}}

\def\Ac{{\cal A}}

\def\eps{\epsilon}

%% file: highstring.bbl
\begin{thebibliography}{10}
\newcommand{\enquote}[1]{``#1''}
\providecommand{\url}[1]{\texttt{#1}}
\providecommand{\urlprefix}{URL }
\providecommand{\eprint}[2][]{\url{#2}}

\bibitem{Gross:1987ar}
D.~J. Gross and P.~F. Mende, \enquote{{String Theory Beyond the Planck Scale},}
  Nucl. Phys. B \textbf{303} (1988) 407--454

\bibitem{Stieberger:2018edy}
S.~Stieberger and T.~R. Taylor, \enquote{{Strings on Celestial Sphere},} Nucl.
  Phys. B \textbf{935} (2018) 388--411, \eprint{1806.05688}

\bibitem{Jiang:2021csc}
H.~Jiang, \enquote{{Celestial OPEs and $w_{1+\infty}$ algebra from worldsheet
  in string theory},} JHEP \textbf{01} (2022) 101, \eprint{2110.04255}

\bibitem{Donnay:2023kvm}
L.~Donnay, G.~Giribet, H.~Gonz\'alez, A.~Puhm and F.~Rojas, \enquote{{Celestial
  open strings at one-loop},} JHEP \textbf{10} (2023) 047, \eprint{2307.03551}

\bibitem{Castiblanco:2024hnq}
L.~Castiblanco, G.~Giribet, G.~Marin and F.~Rojas, \enquote{{Celestial strings:
  Field theory, conformally soft limits, and mapping the~worldsheet onto the
  celestial sphere},} Phys. Rev. D \textbf{110} (2024)(12) 126001,
  \eprint{2405.01643}

\bibitem{Stieberger:2024shv}
S.~Stieberger, T.~R. Taylor and B.~Zhu, \enquote{{Carrollian Amplitudes from
  Strings},} JHEP \textbf{04} (2024) 127, \eprint{2402.14062}

\bibitem{Bockisch:2024bia}
D.~Bockisch, \enquote{{Celestial string integrands \& their expansions},} Nucl.
  Phys. B \textbf{1011} (2025) 116792, \eprint{2408.02609}

\bibitem{Fan:2022vbz}
W.~Fan, A.~Fotopoulos, S.~Stieberger, T.~R. Taylor and B.~Zhu,
  \enquote{{Elements of celestial conformal field theory},} JHEP \textbf{08}
  (2022) 213, \eprint{2202.08288}

\bibitem{Casali:2022fro}
E.~Casali, W.~Melton and A.~Strominger, \enquote{{Celestial amplitudes as
  AdS-Witten diagrams},} JHEP \textbf{11} (2022) 140, \eprint{2204.10249}

\bibitem{Fan:2022kpp}
W.~Fan, A.~Fotopoulos, S.~Stieberger, T.~R. Taylor and B.~Zhu,
  \enquote{{Celestial Yang-Mills amplitudes and D = 4 conformal blocks},} JHEP
  \textbf{09} (2022) 182, \eprint{2206.08979}

\bibitem{deGioia:2022fcn}
L.~P. de~Gioia and A.-M. Raclariu, \enquote{{Eikonal approximation in celestial
  CFT},} JHEP \textbf{03} (2023) 030, \eprint{2206.10547}

\bibitem{Gonzo:2022tjm}
R.~Gonzo, T.~McLoughlin and A.~Puhm, \enquote{{Celestial holography on
  Kerr-Schild backgrounds},} JHEP \textbf{10} (2022) 073, \eprint{2207.13719}

\bibitem{Banerjee:2023rni}
S.~Banerjee, R.~Mandal, A.~Manu and P.~Paul, \enquote{{MHV gluon scattering in
  the massive scalar background and celestial OPE},} JHEP \textbf{10} (2023)
  007, \eprint{2302.10245}

\bibitem{Ball:2023ukj}
A.~Ball, S.~De, A.~Yelleshpur~Srikant and A.~Volovich,
  \enquote{{Scalar-graviton amplitudes and celestial holography},} JHEP
  \textbf{02} (2024) 097, \eprint{2310.00520}

\bibitem{Crawley:2023brz}
E.~Crawley, A.~Guevara, E.~Himwich and A.~Strominger, \enquote{{Self-dual black
  holes in celestial holography},} JHEP \textbf{09} (2023) 109,
  \eprint{2302.06661}

\bibitem{Adamo:2024mqn}
T.~Adamo, W.~Bu, P.~Tourkine and B.~Zhu, \enquote{{Eikonal amplitudes on the
  celestial sphere},} JHEP \textbf{10} (2024) 192, \eprint{2405.15594}

\bibitem{Strominger}
A.~Strominger, \enquote{{Strings 2024 -- The Future of String Theory: 100 Open
  Questions},}   (2024)

\bibitem{Veneziano:1968yb}
G.~Veneziano, \enquote{{Construction of a crossing - symmetric, Regge behaved
  amplitude for linearly rising trajectories},} Nuovo Cim. A \textbf{57} (1968)
  190--197

\bibitem{Gross:1987kza}
D.~J. Gross and P.~F. Mende, \enquote{{The High-Energy Behavior of String
  Scattering Amplitudes},} Phys. Lett. B \textbf{197} (1987) 129--134

\bibitem{Gross:1989ge}
D.~J. Gross and J.~L. Manes, \enquote{{The High-energy Behavior of Open String
  Scattering},} Nucl. Phys. B \textbf{326} (1989) 73--107

\bibitem{Schlotterer:2012ny}
O.~Schlotterer and S.~Stieberger, \enquote{{Motivic Multiple Zeta Values and
  Superstring Amplitudes},} J. Phys. A \textbf{46} (2013) 475401,
  \eprint{1205.1516}

\bibitem{Gradshteyn:1702455}
I.~S. Gradshteyn, I.~M. Ryzhik, D.~Zwillinger and V.~Moll, {Table of integrals,
  series, and products; 8th ed.}, Academic Press, Amsterdam (2015)

\bibitem{Mizera:2019vvs}
S.~Mizera and A.~Pokraka, \enquote{{From Infinity to Four Dimensions: Higher
  Residue Pairings and Feynman Integrals},} JHEP \textbf{02} (2020) 159,
  \eprint{1910.11852}

\bibitem{Aprile:2020luw}
F.~Aprile and P.~Vieira, \enquote{{Large $p$ explorations. From SUGRA to big
  STRINGS in Mellin space},} JHEP \textbf{12} (2020) 206, \eprint{2007.09176}

\bibitem{Lee:2015wwa}
J.-C. Lee and Y.~Yang, \enquote{{Review on High energy String Scattering
  Amplitudes and Symmetries of String Theory},}   (2015), \eprint{1510.03297}

\bibitem{Witten:2013pra}
E.~Witten, \enquote{{The Feynman $i \epsilon$ in String Theory},} JHEP
  \textbf{04} (2015) 055, \eprint{1307.5124}

\bibitem{Yoda:2024pie}
T.~Yoda, \enquote{{Complex saddles and time-delay of Veneziano amplitude},}
  (2024), \eprint{2402.06153}

\bibitem{Eberhardt:2024twy}
L.~Eberhardt and S.~Mizera, \enquote{{Lorentzian contours for tree-level string
  amplitudes},} SciPost Phys. \textbf{17} (2024)(3) 078, \eprint{2403.07051}

\bibitem{Kontsevich2001}
M.~Kontsevich and D.~Zagier, Periods, Springer Berlin Heidelberg, Berlin,
  Heidelberg (2001), pp. 771--808

\bibitem{Stieberger:2013wea}
S.~Stieberger, \enquote{{Closed superstring amplitudes, single-valued multiple
  zeta values and the Deligne associator},} J. Phys. A \textbf{47} (2014)
  155401, \eprint{1310.3259}

\bibitem{Stieberger:2014hba}
S.~Stieberger and T.~R. Taylor, \enquote{{Closed String Amplitudes as
  Single-Valued Open String Amplitudes},} Nucl. Phys. B \textbf{881} (2014)
  269--287, \eprint{1401.1218}

\bibitem{Brown:2013gia}
F.~Brown, \enquote{{Single-valued Motivic Periods and Multiple Zeta Values},}
  SIGMA \textbf{2} (2014) e25, \eprint{1309.5309}

\bibitem{Brown:2018omk}
F.~Brown and C.~Dupont, \enquote{{Single-valued integration and double copy},}
  J. Reine Angew. Math. \textbf{2021} (2021)(775) 145--196, \eprint{1810.07682}

\bibitem{Brown:2019wna}
F.~Brown and C.~Dupont, \enquote{{Single-valued integration and superstring
  amplitudes in genus zero},} Commun. Math. Phys. \textbf{382} (2021)(2)
  815--874, \eprint{1910.01107}

\bibitem{Kawai:1985xq}
H.~Kawai, D.~C. Lewellen and S.~H.~H. Tye, \enquote{{A Relation Between Tree
  Amplitudes of Closed and Open Strings},} Nucl. Phys. B \textbf{269} (1986)
  1--23

\bibitem{Beisert:2006ez}
N.~Beisert, B.~Eden and M.~Staudacher, \enquote{{Transcendentality and
  Crossing},} J. Stat. Mech. \textbf{0701} (2007) P01021,
  \eprint{hep-th/0610251}

\bibitem{Mizera:2019gea}
S.~Mizera, \enquote{{Aspects of Scattering Amplitudes and Moduli Space
  Localization},} Ph.D. thesis, Princeton, Inst. Advanced Study (2020),
  \eprint{1906.02099}

\bibitem{Mansour:2015awy}
T.~Mansour and M.~Schork, {Commutation Relations, Normal Ordering, and Stirling
  Numbers}, Chapman and Hall/CRC (2015)

\bibitem{Pasterski:2016qvg}
S.~Pasterski, S.-H. Shao and A.~Strominger, \enquote{{Flat Space Amplitudes and
  Conformal Symmetry of the Celestial Sphere},} Phys. Rev. D \textbf{96}
  (2017)(6) 065026, \eprint{1701.00049}

\bibitem{Pasterski:2017ylz}
S.~Pasterski, S.-H. Shao and A.~Strominger, \enquote{{Gluon Amplitudes as 2d
  Conformal Correlators},} Phys. Rev. D \textbf{96} (2017)(8) 085006,
  \eprint{1706.03917}

\bibitem{Donnay:2018neh}
L.~Donnay, A.~Puhm and A.~Strominger, \enquote{{Conformally Soft Photons and
  Gravitons},} JHEP \textbf{01} (2019) 184, \eprint{1810.05219}

\bibitem{Whittaker_Watson_1996}
E.~T. Whittaker and G.~N. Watson, A Course of Modern Analysis, Cambridge
  Mathematical Library, 4 edition, Cambridge University Press (1996)

\bibitem{Arkani-Hamed:2020gyp}
N.~Arkani-Hamed, M.~Pate, A.-M. Raclariu and A.~Strominger, \enquote{{Celestial
  amplitudes from UV to IR},} JHEP \textbf{08} (2021) 062, \eprint{2012.04208}

\bibitem{Stieberger:2016xhs}
S.~Stieberger, \enquote{{Periods and Superstring Amplitudes},}   (2016),
  \eprint{1605.03630}

\bibitem{Mizera:2022sln}
S.~Mizera and S.~Pasterski, \enquote{{Celestial geometry},} JHEP \textbf{09}
  (2022) 045, \eprint{2204.02505}

\bibitem{Dvali:2014ila}
G.~Dvali, C.~Gomez, R.~S. Isermann, D.~L\"ust and S.~Stieberger,
  \enquote{{Black hole formation and classicalization in ultra--Planckian
  $2\rightarrow N$ scattering},} Nucl. Phys. B \textbf{893} (2015) 187--235,
  \eprint{1409.7405}

\bibitem{Schild:1976vq}
A.~Schild, \enquote{{Classical Null Strings},} Phys. Rev. D \textbf{16} (1977)
  1722

\bibitem{Bagchi:2015nca}
A.~Bagchi, S.~Chakrabortty and P.~Parekh, \enquote{{Tensionless Strings from
  Worldsheet Symmetries},} JHEP \textbf{01} (2016) 158, \eprint{1507.04361}

\bibitem{Bagchi:2013bga}
A.~Bagchi, \enquote{{Tensionless Strings and Galilean Conformal Algebra},} JHEP
  \textbf{05} (2013) 141, \eprint{1303.0291}

\bibitem{Bagchi:2024qsb}
A.~Bagchi, P.~Chakraborty, S.~Chakrabortty, S.~Fredenhagen, D.~Grumiller and
  P.~Pandit, \enquote{{Boundary Carrollian Conformal Field Theories and Open
  Null Strings},} Phys. Rev. Lett. \textbf{134} (2025)(7) 071604,
  \eprint{2409.01094}

\bibitem{Francia:2007qt}
D.~Francia, J.~Mourad and A.~Sagnotti, \enquote{{Current Exchanges and
  Unconstrained Higher Spins},} Nucl. Phys. B \textbf{773} (2007) 203--237,
  \eprint{hep-th/0701163}

\bibitem{Bagchi:2019cay}
A.~Bagchi, A.~Banerjee and P.~Parekh, \enquote{{Tensionless Path from Closed to
  Open Strings},} Phys. Rev. Lett. \textbf{123} (2019)(11) 111601,
  \eprint{1905.11732}

\bibitem{Alday:2023pzu}
L.~F. Alday, T.~Hansen and M.~Nocchi, \enquote{{High Energy String Scattering
  in AdS},} JHEP \textbf{02} (2024) 089, \eprint{2312.02261}

\bibitem{Alday:2023jdk}
L.~F. Alday, T.~Hansen and J.~A. Silva, \enquote{{Emergent Worldsheet for the
  AdS Virasoro-Shapiro Amplitude},} Phys. Rev. Lett. \textbf{131} (2023)(16)
  161603, \eprint{2305.03593}

\bibitem{miller_applied_2006}
P.~Miller, Applied Asymptotic Analysis, volume~75 of Graduate Studies in
  Mathematics, American Mathematical Society

\bibitem{Bender1979AdvancedMM}
C.~M. Bender and S.~A. Orszag, \enquote{Advanced Mathematical Methods For
  Scientists And Engineers: Asymptotic Methods and Perturbation Theory,}
  (1999)

\bibitem{Schafer}
R.~Schafer and R.~Kouyoumjian, \enquote{Higher order terms in the saddle point
  approximation,} Proceedings of the IEEE \textbf{55} (1967)(8) 1496--1497

\bibitem{Mirkov}
N.~Fabiano and N.~Mirkov, \enquote{Saddle point approximation to Higher order,}
  Vojnotehnicki glasnik \textbf{70} (2022) 447--460

\end{thebibliography}
